\pdfoutput=1
\documentclass[nofootinbib, twocolumn, aps, prd, superscriptaddress]{revtex4-1}

\usepackage{longtable}
\usepackage{array}
\usepackage{amsmath}
\usepackage{amsfonts}
\newcolumntype{C}[1]{>{\centering}m{#1}}

\def\apj{{ApJ}}                 
\def\aap{{A\&A}}                

\def\mnras{Mon.~Not.~Roy.~Astron.~Soc.}             
\def\prd{{Phys.~Rev.~D}}        
\usepackage{graphicx}
\usepackage{natbib}
\usepackage{color}
\newcommand{\beq}{\begin{equation}}
\newcommand{\beqa}{\begin{eqnarray}}
\newcommand{\eeq}{\end{equation}}
\newcommand{\eeqa}{\end{eqnarray}}

\newcommand{\simgt}{\lower.5ex\hbox{$\; \buildrel > \over \sim \;$}}
\newcommand{\simlt}{\lower.5ex\hbox{$\; \buildrel < \over \sim \;$}}

\newcommand{\bd}[1]{\mbox{\boldmath $#1$}}

\usepackage[usenames,dvipsnames]{xcolor}
\newcommand{\change}[1]{{\color{black} #1}}

\begin{document}


\title{Denoising Weak Lensing Mass Maps with Deep Learning}

\author{Masato Shirasaki}
 \email{masato.shirasaki@nao.ac.jp}
\affiliation{%
 National Astronomical Observatory of Japan (NAOJ), Mitaka, Tokyo 181-8588, Japan
}%


\author{Naoki Yoshida}
\affiliation{
 Department of Physics, University of Tokyo, Tokyo 113-0033, Japan
}%
\affiliation{
 Kavli Institute for the Physics and Mathematics of the Universe (WPI), University of Tokyo, Kashiwa, Chiba 277-8583, Japan
}%
\affiliation{
Institute for Physics of Intelligence, University of Tokyo, Tokyo 113-0033, Japan
}
\author{
Shiro Ikeda
}
\affiliation{%
The Institute of Statistical Mathematics, 10-3 Midori-cho, Tachikawa, Tokyo 190-8562, Japan
}%
\affiliation{Department of Statistical Science,
Graduate University for Advanced Studies, 10-3 Midori-cho, Tachikawa, Tokyo 190-8562, Japan
}


\date{\today}

\begin{abstract}
Weak gravitational lensing is a powerful probe
of the large-scale cosmic matter distribution. Wide-field galaxy surveys
allow us to generate the so-called weak lensing maps, but actual observations 
suffer from noise due to imperfect measurement of galaxy shape distortions and to the limited number density of 
the source galaxies. In this paper, we explore a deep-learning approach to reduce the noise.
We develop an image-to-image translation method with conditional adversarial networks (CANs), 
which learn efficient mapping from an input noisy weak lensing map to the underlying noise field. 
We train the CANs using $30000$ image pairs obtained from $1000$ ray-tracing simulations of weak gravitational lensing. 
We show that the trained CANs reproduce the true one-point probability distribution function (PDF)
of the noiseless lensing map with a bias less than $1\sigma$ on average, where $\sigma$ is the statistical error.
We perform a Fisher analysis to make forecast for cosmological parameter inference with the one-point lensing PDF.
By our denoising method using CANs, the first derivative of the PDF with respect to
the cosmic mean matter density and the amplitude of the primordial curvature perturbations becomes larger by $\sim50\%$. This allows us to improve the cosmological constraints by $\sim30-40\%$ with using observational data from ongoing and upcoming galaxy imaging surveys.
\end{abstract}

\pacs{Valid PACS appear here}
\maketitle


\section{Introduction}

Gravitational lensing is a relativistic effect that causes characteristic distortion of the images of distant astrophysical sources.
The degree of distortion is determined by the gravitational potential of intervening mass (lens) and 
the geometry between the lens and source objects \cite{Bartelmann:1999yn}.
Although the induced distortion is tiny for individual sources, 
averaging over a large number of sources 
can reveal the gravitational lensing effect 
due to large-scale mass distribution in the Universe.
In the literature, the lensing effect caused by the large-scale structure is referred to as weak lensing effect.
The great advantage of measuring the weak lensing effect is that it enables us to study the matter density distribution 
in a physical and unbiased manner in principle. 
Statistical analysis of weak lensing effect is one of the most important studies
in modern cosmology; one can extract rich information of gravitational clumping of dark matter and cosmic expansion
(see, Refs~\cite{2009NuPhS.194...76H,2010GReGr..42.2177H,Kilbinger:2014cea} for reviews).

The two-point angular correlation function of 
shapes of sources, or its Fourier-space counterpart known as power spectrum,
are commonly used to characterize weak lensing maps.
Although the power spectrum provides a complete statistical description of a random Gaussian field, 
numerical simulations of weak lensing effect have shown that weak lensing 
maps have non-Gaussian properties.
Hence the power spectrum alone cannot fully describe a weak lensing map \cite{Jain:1999ir, White:1999xa, Hamana:2001vz}.
Various statistical methods have been proposed to study the non-Gaussian features
\cite{Matsubara:2000dg, Sato:2001cb, Zaldarriaga:2002qt, Takada:2002hh, Pen:2003vw, Jarvis:2003wq, Wang:2008hi, 2010MNRAS.402.1049D, 2010PhRvD..81d3519K, 2010ApJ...719.1408F, Shirasaki:2013zpa, 2015A&A...576A..24L, 2015PhRvD..91j3511P, 2018arXiv181002374C, Schmelzle:2017vwd, Gupta:2018eev, Ribli:2018kwb}.
Most of these proposed methods consider two-dimensional maps, i.e., images.

Distortion of galaxy shape is commonly used as a measure of weak lensing effect.
The weak lensing effect of individual galaxies is expected to be much smaller than the intrinsic shape in practice \cite{Kilbinger:2014cea},
and the shape measurement is affected by various observational effects \cite{Mandelbaum:2013esa}, 
rendering the measurement and the resulting statistics uncertain.
These are altogether called "shape noises", which make weak lensing maps estimated from galaxy shape 
to be intrinsically noisy. Often the original cosmological information imprinted in the map is obscured.
There exist several approaches to reduce the shape noise in weak lensing maps 
in the literature \cite{Bartelmann:1996tj, Bridle:1998ee, Marshall:2001ax, Starck:2005ek, 2012A&A...540A..34D, Jeffrey:2018cvw}, but it is still challenging to obtain a completely noiseless 
weak lensing map from observations with an angular resolution of $\sim1$ arcmin.
Noise reduction in weak lensing maps would be crucial for various cosmological analyses 
including robust search for clusters of galaxies \cite{Hamana:2003ts, 2012MNRAS.425.2287H, 2015MNRAS.453.3043S, 2016MNRAS.459.2762H}
and detection of diffuse components such as filaments in a cosmic mass density field \cite{2014MNRAS.441..745H}.
It is also known that non-Gaussianity of noiseless lensing maps can be utilized to avoid the degeneracy of cosmological parameters effectively,
and hence to allow precise measurement of time evolution in dark energy density (e.g., Ref.~\cite{2012PhRvD..85j3513K}) and neutrino masses (e.g., Refs.~\cite{2018arXiv181002374C,2018arXiv181208206M, 2019PhRvD..99f3527L})

In this paper, we explore a deep-learning approach to 
generate a high-resolution weak lensing map from noisy, observed one.
The procedure is essentially an image-to-image translation from a noisy input map to the underlying noise field.
We adopt the conditional adversarial networks for image-to-image translation used in Ref.~\cite{2016arXiv161107004I}.
We train and validate the networks by using 30000 noisy weak lensing maps based on cosmological weak lensing simulations.
The training is performed with effective angular resolution of $\sim1.5$ arcmin,
and we assume that the root-mean-square error of the input noisy maps is totally dominated by the shape noise.
We then test the trained networks by using additional 1000 data sets and investigate
if our deep-learning method can reproduce two summary statistics of weak lensing maps in absence of noise:
one is the power spectrum and the other is one-point probability distribution function (PDF).
We also aim at reconstruction of noise-free weak lensing map on pixel-by-pixel basis.
We finally examine if our denoising method is a valid approach for two common weak-lensing analyses of
detection of massive galaxy clusters and precise determination of cosmological model.

The rest of the present paper is organized as follows. 
In Section~\ref{sec:WL}, 
we summarize the basics of gravitational lensing. 
In Section~\ref{sec:method}, we describe the conditional adversarial networks adopted in this paper 
and provide how to produce the data set for training, validating, and testing the networks. 
In Section~\ref{sec:results}, we show the performance of our trained networks when applying them to test data set.
We also study the applicability of our denoising method in cosmological analyses in Section~\ref{sec:Fisher}.
Future prospects in upcoming surveys with a higher source number density are discussed in Section~\ref{sec:future}.
Concluding remarks and discussions are given in Section~\ref{sec:con}. 

\section{Weak gravitational lensing}\label{sec:WL}
\subsection{Basics}\label{subsec:basics}

Weak lensing effect is commonly characterized by
the distortion of image of a source object (galaxy) by the following $2\times2$ matrix
between the observed position of a source object $\bd{\theta}_{\rm obs}$ and the true (unlensed) position $\bd{\theta}_{\rm true}$:
\beqa
A_{ij} = \frac{\partial \theta_{\rm true}^{i}}{\partial \theta_{\rm obs}^{j}}
           \equiv \left(
\begin{array}{cc}
1-\kappa -\gamma_{1} & -\gamma_{2}-\omega  \\
-\gamma_{2}+\omega & 1-\kappa+\gamma_{1} \\
\end{array}
\right), \label{eq:distortion_tensor}
\eeqa
where $\kappa$ is the convergence, $\gamma$ is the shear, and $\omega$ is the rotation.
In the weak lensing regime ($\kappa, \gamma \ll 1$), the convergence
can be expressed as the integral of the density contrast of underlying matter density field $\delta_{\rm m}(\bd{x})$ 
with a weight over redshift \cite{Bartelmann:1999yn},
\beqa
\kappa(\bd{\theta}) &=& \int_{0}^{\infty}{\rm d}\chi\, W_{\kappa}(\chi) \delta_{\rm m}(r(\chi)\bd{\theta}, \chi), \\ \label{eq:delta2kappa}
W_{\kappa}(\chi) &=& \frac{3}{2}\left(\frac{H_{0}}{c}\right)^2\Omega_{\rm m0}(1+z(\chi))r(\chi) \nonumber \\
&&
\,\,\,\,\,
\,\,\,\,\,
\,\,\,\,\,
\,\,\,\,\,
\times
\int_{\chi}^{\infty}{\rm d}\chi^{\prime}\,
p(\chi^{\prime})\frac{r(\chi^{\prime}-\chi)}{r(\chi^{\prime})},
\label{eq:lens_kernel}
\eeqa
where 
$H_{0}$ is the present-day Hubble constant,
$\Omega_{\rm m0}$ is the matter density parameter at present, 
$\chi(z)$ is the radial comoving distance to redshift $z$,
$r(\chi)$ is the angular diameter distance, and $p(\chi)$ represents 
the source distribution normalized to $\int{\rm d}\chi\,p(\chi)=1$.
Throughout this paper, 
we assume that the source galaxies are located at a single plane at redshift of $z_{\rm source} = 1$, 
i.e. $p(\chi)=\delta(\chi-\chi_{1})$ where $\delta(x)$ is the Dirac delta function and 
$\chi_1=\chi(z=1)$ for simplicity.

\subsection{Estimator of convergence field}\label{subsec:est_kappa}

The lensing convergence $\kappa$ is of our primary interest 
since it contains rich cosmological information.
In the following, we summarize how to estimate $\kappa$ from observables in modern galaxy imaging surveys.
We first define the smoothed convergence map (field) as
\beqa
\hat{\kappa}(\bd{\theta}) = 
\int {\rm d}^2 \bd{\phi}\, \kappa(\bd{\theta}-\bd{\phi}) 
U (\bd{\phi}),
\eeqa
where $U$ is the filter function to be specified below.
We can calculate the same quantity by smoothing the shear field $\gamma$ as
\beqa
\hat{\kappa} (\bd{\theta}) = \int {\rm d}^2 \bd{\phi} \ \gamma_{+}(\bd{\phi}:\bd{\theta}) Q_{+}(\bd{\phi}), \label{eq:ksm}
\eeqa
where $\gamma_{+}$ is the tangential component of the shear 
at position $\bd{\phi}$ relative to the point $\bd{\theta}$.
The filter function for the shear field $Q_{+}$ is related to $U$ by
\beqa
Q_{+}(\theta) = \int_{0}^{\theta} {\rm d}\theta^{\prime} \ \theta^{\prime} U(\theta^{\prime}) - U(\theta).
\label{eq:U_Q_fil}
\eeqa
We consider a filter function $Q_{+}$ that has a finite extent.
In such cases, one can write
\beqa
U(\theta) = 2\int_{\theta}^{\theta_{o}} {\rm d}\theta^{\prime} \ \frac{Q_{+}(\theta^{\prime})}{\theta^{\prime}} - Q_{+}(\theta),
\eeqa
where $\theta_{o}$ is the outer boundary of the filter function.
Note that the filter function $U$ should be compensated 
because the smoothed field $\hat{\kappa}$ does not depend on 
undetermined constant \cite{Schneider:1996ug}.

In this paper, we consider the truncated Gaussian filter (for $U$):
\beqa
U(\theta) &=& \frac{1}{\pi \theta_{G}^{2}} 
\exp \left( -\frac{\theta^2}{\theta_{G}^2} \right) \nonumber \\
&&
\,\,\,\,\,\,\,\,
\,\,\,\,\,\,\,\,
\,\,\,\,\,\,\,\,
-\frac{1}{\pi \theta_{o}^2}\left[ 1-\exp \left(-\frac{\theta_{o}^2}{\theta_{G}^2} \right) \right], \label{eq:filter_kappa}\\
Q_{+}(\theta) &=& \frac{1}{\pi \theta^{2}}\left[ 1-\left(1+\frac{\theta^2}{\theta_{G}^2}\right)\exp\left(-\frac{\theta^2}{\theta_{G}^2}\right)\right],
\label{eq:filter_gamma}
\eeqa
for $\theta \leq \theta_{o}$ and $U = Q_{+} = 0$ elsewhere.
Throughout this paper, we set $\theta_{o} = 30$ arcmin
and $\theta_{G} = 1.5$ arcmin.
The choice of $\theta_{G} = 1.5$ arcmin 
is found to be an optimal smoothing scale for 
the detection of massive galaxy clusters using weak lensing for 
$z_{\rm source}\sim1$ \cite{2012MNRAS.425.2287H, 2015MNRAS.453.3043S}.

In actual observations, ellipticity of galaxies is used as an indicator of the shear field.
In the weak-field limit ($\kappa, \gamma \ll 1$), 
the observed ellipticity of galaxy can be decomposed into two parts as
\beqa
\epsilon_{\rm obs} = \epsilon_{\rm N} +  \gamma,
\eeqa
where $\epsilon_{\rm obs}$ and $\epsilon_{\rm N}$ are the observed ellipticity and shape noise, respectively.
In typical galaxy imaging surveys, the shape noise term is mostly contributed by the intrinsic ellipticity of source galaxies 
and the shape measurement inaccuracy. Since we expect both are independent of weak lensing shear $\gamma$,
the smoothed convergence estimated by Eq.~(\ref{eq:ksm}) can be expressed as
\beqa
\kappa_{\rm obs} = \kappa_{\rm N} + \kappa,
\eeqa
where the left-hand side represents the observed convergence, while $\kappa_{\rm N}$ in the right-hand side 
is the noise convergence from the shape noise.
The primary purpose of this paper is to estimate the underlying convergence field $\kappa$ from the observed map $\kappa_{\rm obs}$.
To this end, we adopt an image-to-image translation based on conditional adversarial networks.

\section{Method}\label{sec:method}

\subsection{Image-to-Image Translation}

We use conditional adversarial networks developed in Ref~\cite{2016arXiv161107004I}, 
referred to as {\tt pix2pix}, for denoising the weak lensing convergence field.
{\tt pix2pix} is designed so as to learn mapping from input to output, but also to learn a loss function associated with this mapping. Therefore, it provides a generic approach to image-to-image translation problems, and is free from the formulation of loss function in the networks on a problem-by-problem basis.

The networks have two main pieces, a generator and a discriminator. 
The generator applies some transform to the input image to get an output image. 
The discriminator compares the input image to an unknown image 
(either a target image from the data set or an output image from the generator) and 
tries to guess if it is produced by the generator.

The generator in {\tt pix2pix} uses a U-Net which is a kind of encoder-decoder structure \cite{2015arXiv150504597R}.
On the encoder part in the generator, an input image is progressively compressed with eight convolution layers.
In this compression process, the generator tries to learn the important features of the input at different scales.
Each convolution layer consists of convolution with a kernel size of $5\times5$,
the batch normalization, and the application of activation function of leaky ReLU with a leak slope of 0.2.
When arriving at the final convolution layer, the
generator
performs the inverse operation of convolution layer 
and combines the simplified feature from the final convolution layer into more and more complicated representations.
In addition, the U-Net has additional skip connections between mirrored layers in the encoder and decoder stacks,
allowing to propagate the small-scale information that would be lost as the size of the images decreases in the encoder.

In the discriminator, a given set of target and input images are reduced with 4 convolution layers and then all responses are averaged out to provide the ultimate output. The final output in the discriminator is used to determine if the input is real or fake.
Note the target image in the discriminator can be randomly selected from the ground truth or the
output of the generator.

\begin{figure*}
\begin{center}
       \includegraphics[clip, width=1.5\columnwidth, viewport=5 50 1010 768]
       {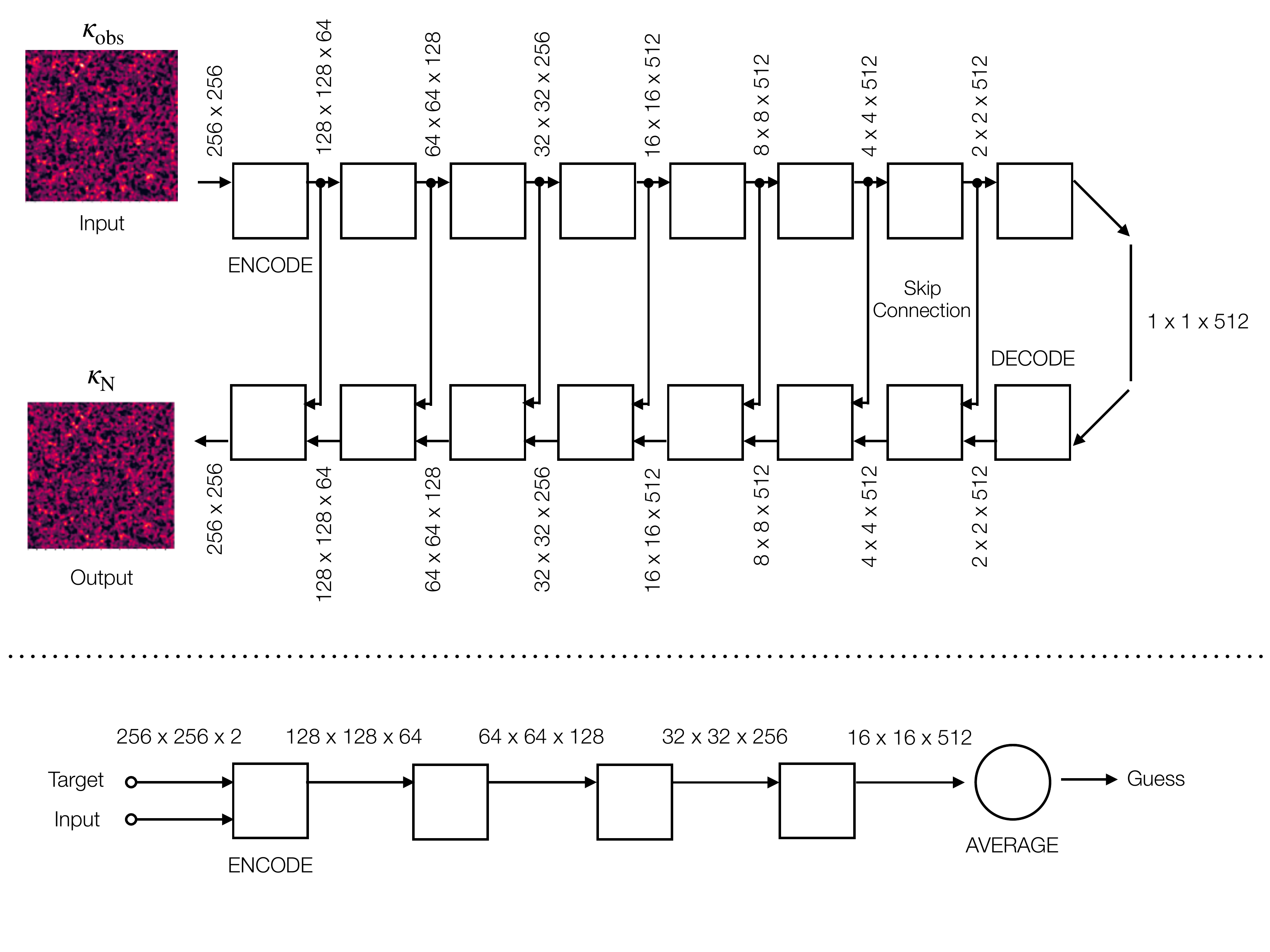}
     \caption{
     The structure of conditional adversarial networks in this paper. The upper portion represents the generator
     that tries to translate an input noisy weak lensing map $\kappa_{\rm obs}$ to the underlying noise field $\kappa_{\rm N}$.
     The lower shows the discriminator which guesses if the input image has been produced by the generator or not.
     Both of the generator and discriminator consist of a series of (de)convolution layers and the resulting number of model parameters in the networks is close to 400000.
     \label{fig:pix2pix}
  } 
    \end{center}
\end{figure*}

In this paper, we adopt the structure of the generator and the discriminator as shown in Figure~\ref{fig:pix2pix}.
To obtain underlying convergence from an input noisy map, 
we first perform an image-to-image translation from a noisy lensing map $\kappa_{\rm obs}$ to the underlying noise field $\kappa_{\rm N}$\footnote{
In Appendix~\ref{app:train_kappa}, we also examine the training so that the generator can predict the underlying noiseless convergence field from noisy input, but such training is found to be less effective for denoising.
}.
Once the noise field is predicted by the networks, 
we then estimate the underlying convergence $\kappa$ by subtracting the predicted noise 
from the observed map $\kappa_{\rm obs}$.

\subsection{Ray-tracing simulation}\label{subsec:sims}

To construct a large data set for deep learning, we utilize ray-tracing simulations of gravitational lensing
in Ref~\cite{Sato:2009ct}. The simulations are based on 
$200$ realizations of high-resolution $N$-body simulations with a box size of 240 $h^{-1}\, {\rm Mpc}$ 
on a side. The $N$-body simulations are performed with
the number of particles of $256^3$ for the concordance $\Lambda$CDM  model\footnote{The following cosmological parameters 
are adopted in the simulations:
the matter density $\Omega_{\rm m0}=0.238$, 
the baryon density $\Omega_{b} = 0.042$,
the dark energy density $\Omega_{\Lambda}=1-\Omega_{\rm m0}=0.762$,
the equation of state parameters of dark energy $w=-1$,
the scalar spectral index $n_s = 0.958$,
the amplitude of curvature perturbations
$A_s=2.35\times10^{-9}$ at $k=0.002\, {\rm Mpc}^{-1}$,
Hubble parameter $h=0.732$,
and the variance of the present-day density
fluctuation in a sphere of radius $8 \, h^{-1}\, {\rm Mpc}$
$\sigma_8 = 0.76$.}.
The mass of $N$-body particle is set to $5.3\times10^{10}\, h^{-1}M_{\odot}$.
We work with a single source redshift at
$z_{\rm source}=1$.
The ray-tracing simulations are performed on $2048^2$ pixels with the pixel size of 0.15 arcmin.
From the 200 $N$-body simulations,
they have produced 1000 realizations of $5\times5$ squared-degrees lensing fields by rotating and shifting
the structures in $N$-body simulations randomly. Details of the ray-tracing simulations are found in Ref~\cite{Sato:2009ct}.

When producing a training set for deep learning, we first degrade the lensing shear fields on $512^2$ pixels 
by averaging over nearby 4 by 4 pixels (the degraded pixel size is then 0.6 arcmin).
We then randomly subtract the contiguous area consisting of $256^2$ pixels to increase the number of training sets.
We perform the random subtraction 100 times for individual simulations with $512^2$ pixels.
For the main training data set, we use 600 realizations of lensing shear among 1000 parent realizations, and then have 
60000 shear fields with a total sky coverage of $2.5\times2.5$ squared degrees.
We also produce another 1000 shear fields for testing the networks from remaining 10 realizations.
To create a noisy field, we include galaxy shape noise $\epsilon_{\rm N}$ in our simulation 
by adding to random ellipticities which follow the two-dimensional Gaussian distribution as
\beqa
P(\epsilon_{\rm N}) = \frac{1}{\pi \sigma_{\rm N}^2} \exp\left(-\frac{\epsilon_{\rm N}^2}{\sigma_{\rm N}^2}\right), \label{eq:noise}
\eeqa
where $\epsilon_{\rm N} = \sqrt{\epsilon_{{\rm N}, 1}^2+\epsilon_{{\rm N}, 2}^2}$ and 
$\sigma_{\rm N}^2= \sigma_{\epsilon}^2/(n_{\rm gal}\theta^2_{\rm pix})$
with the pixel size of $\theta_{\rm pix}=0.6$ arcmin.
In this paper, we set $\sigma_{\epsilon}=0.35$ and $n_{\rm gal}=20$ ${\rm arcmin}^{-2}$.
These correspond to the typical value of the current-generation ground-based imaging survey \cite{Mandelbaum:2017dvy}.
We prepare 60000 and 1000 realizations of the random ellipticities for training and testing, respectively.

For a given data of shear, we perform the smoothing as in Section.~\ref{subsec:est_kappa} to produce a smoothed convergence field. Note that we produce the smoothed fields from noiseless shear as well as noisy shear on realization-by-realization basis.

\subsection{Setup of Training and Validation}\label{subsec:training}

The objective of our networks is expressed as
\beq
{\rm arg}\, {\rm min}_{G}\, {\rm max}_{D}\,
\Big\{{\cal L}_{\rm cGAN}(G, D) + \lambda {\cal L}_{\rm L1}\Big\},
\label{eq:obs_pix2pix}
\eeq
where $G$ is the generator and $D$ is the discriminator.
We here introduce two loss functions as 
\beqa
{\cal L}_{\rm cGAN}(G, D) &=& \mathbb{E}_{x,y}\log D(x, y) \nonumber \\
&&
\,\,\,
+\mathbb{E}_{x,z}\log\left\{1-D(x, G(x,z))\right\}, \\
{\cal L}_{\rm L1}(G) &=& \mathbb{E}_{x,y,z}\, \sum_{\rm map} \left|y-G(x, z)\right|, \label{eq:L1norm}
\eeqa
where $x$ is the input vector, $y$ is the output vector, and $z$ is a random noise vector at the bottom layer of the generator.
In Eq.~(\ref{eq:L1norm}), the summation runs over all  the pixels in a map.
In the training, we alternate between one gradient descent step  on $D$, then one step on $G$. 
As suggested in Ref.~\cite{2014arXiv1406.2661G}, we train to maximize
the term of $\log D(x,G(x,z))$. In addition,  
we divide the objective by 2 while optimizing $D$, 
which slows down the learning rate of $D$ relative to $G$. 

When training the networks, we use the minibatch Stochastic Gradient Descent (SGD) method and apply the Adam solver \cite{2014arXiv1412.6980K}, 
with learning rate $0.0002$, momentum parameters $\beta_1=0.5$ and $\beta_2=0.9999$.
We also set $\lambda=100$ in Eq.~(\ref{eq:obs_pix2pix}).
All the networks in this paper are trained with a batch size of 1. 
We initialize the model parameters in the networks from a Gaussian  distribution with mean $0$ and standard deviation $0.02$.
We train our networks using the TensorFlow implementation\footnote{\url{https://github.com/yenchenlin/pix2pix-tensorflow}}
of {\tt pix2pix} on a single NVIDIA Quadro P5000 GPU.
While processing, we randomly select training and validation data from the input data sets.
Each network is validated every time it learns 100 image pairs. 
The training procedure takes $0.5-2$ seconds per an image pair, and about a few hours for learning 10000 image pairs.

\begin{figure}
\begin{center}
       \includegraphics[clip, width=0.95\columnwidth]
       {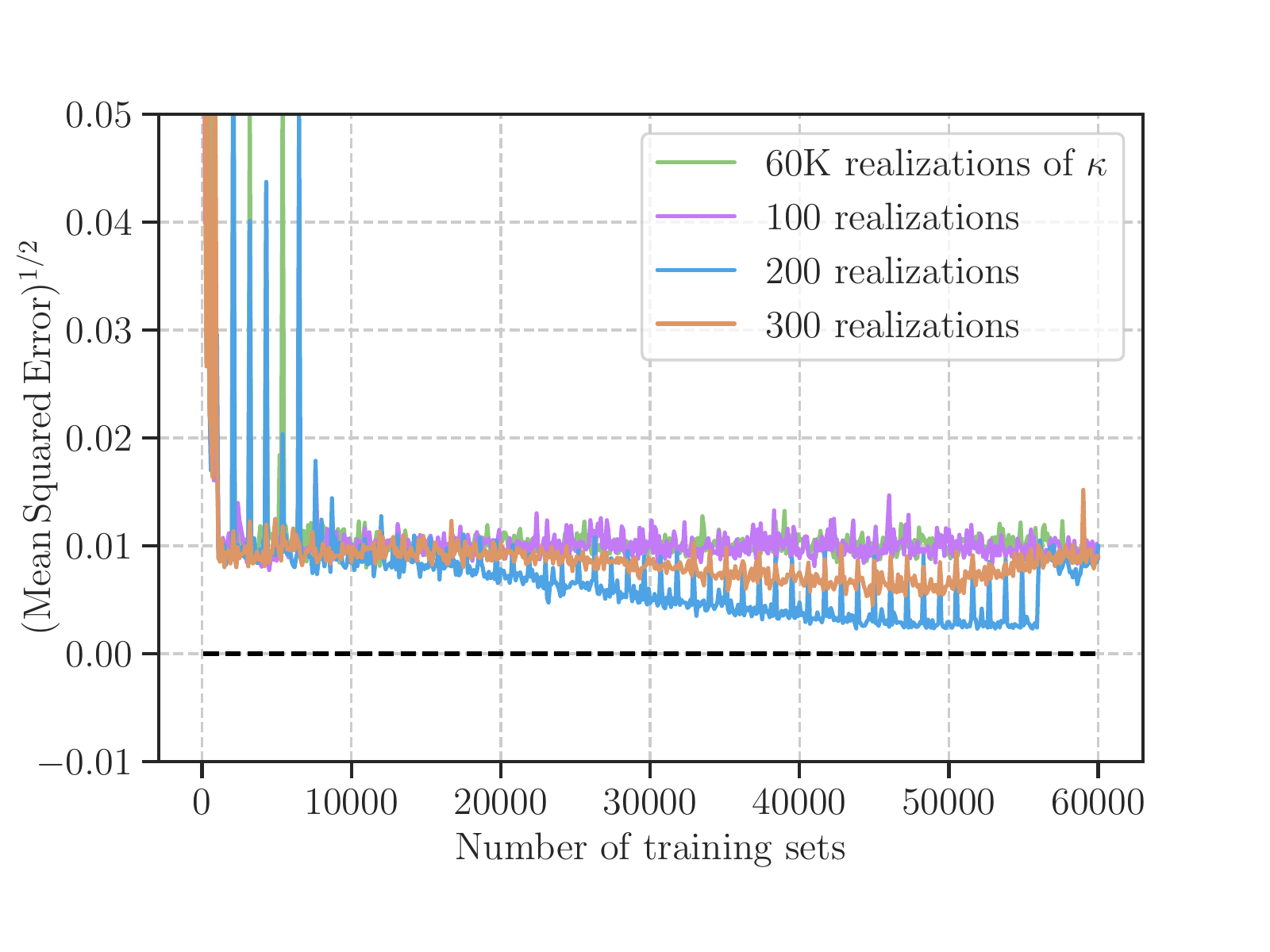}
     \caption{
     The mean squared error of the lensing convergence maps in the training process.
     We measure the mean squared error of the validation data set at different epochs in training.
     In this figure, the colored lines show the training processes based on different realizations of the true convergence:
     {\it Green}: the number of realization of true maps is 60000.
     {\it Purple}: Among 60000 realizations (green), we select 100 realizations before training.
     {\it Blue}: Similar to purple one, but we select 200 realizations.
     {\it Orange}: Similar to purple one, but we select 300 realizations.
     This figure shows that $\simeq200$ realizations of noiseless convergence is suitable to learn the image-to-image translation in
     an efficient way.
     \label{fig:MSE}
  } 
    \end{center}
\end{figure}

Producing a realization of noisy data set $\kappa_{\rm obs}$ is non-trivial for our training task, 
because we use two different independent realizations of the noise $\kappa_{\rm N}$ and lensing field of interest, $\kappa$. 
To optimize the training, we test four cases as:
\begin{enumerate}
    \item The input data set consists of 60000 independent realizations of $\kappa_{\rm N}$. When adding the lensing convergence $\kappa$ to produce a noisy data, we use different realizations of $\kappa$ from 60000 different realizations.
    \item Similar to case 1, but we restrict the number of realization of $\kappa$ to be 300. When adding $\kappa$ to the noise $\kappa_{\rm N}$,
    we choose the map from 300 realizations at random.
    \item Similar to case 2, but we use the number of realization of $\kappa$ to be 200.
    \item Similar to case 3, but we use the number of realization of $\kappa$ to be 100.
\end{enumerate}
In Case 1, we train the networks using 60000 independent noisy maps.
We expect that there is some optimal number of realizations of $\kappa$ in the training process, 
since too many realizations will make an efficient learning difficult for the networks, 
while fewer realizations can not provide a sufficient information for the image-to-image translation.
From this point of view, Case 1 corresponds to oversampling of $\kappa$ for training.

To find the optimal number of realizations of $\kappa$, we measure the mean squared error (MSE) in image space for validation sets.
Figure~\ref{fig:MSE} shows the MSEs for four different input data sets as a function of the number of training sets.
Note that the figure illustrates the proficiency level of networks for denoising, not the convergence check in the training process.
Clearly, efficient learning of image-to-image translation is achieved when we use $\sim200$ realizations of $\kappa$ 
to produce noisy data set. In the following, we use 200 realizations
to produce noisy data for training.
The predicted noise images by the generator with $\simgt40000$ training sets show unrealistic fluctuations 
between nearby pixels, while the predicted noise images are smooth enough if the network uses $20000-40000$ training sets.

In general, the optimal number of training sets depends on a number of factors.
For our purpose of cosmology study here, it is important to train so that the estimation of summary statistics
is unbiased.
Hence, we try to find the optimal number by using two common summary statistics in cosmological studies.
One is power spectrum defined as
\beqa
\langle \tilde{\kappa}(\bd{\ell}_{1}) \tilde{\kappa}(\bd{\ell}_{2})\rangle \equiv (2\pi)^2 \delta^{(2)}(\bd{\ell}_{1}-\bd{\ell}_{2})C_{\kappa}(\ell_1),
\eeqa
where $\tilde{\kappa}$ is the Fourier transform of a noiseless convergence field, 
$\delta^{(n)}(\bd{x})$ is the Delta function in $n$-dimensional space,
and $C_{\kappa}(\ell_1)$ is the power spectrum. 
The other statistic is the one-point probability distribution function (PDF) ${\cal P}$ of lensing convergence in real space.
Note that the power spectrum is a measure of Gaussian information in $\kappa$ fields,
while the PDF can capture some non-Gaussian information such as skewness and kurtosis.

\begin{figure*}
\begin{center}
       \includegraphics[clip, width=2.2\columnwidth, viewport=100 50 1440 576]
       {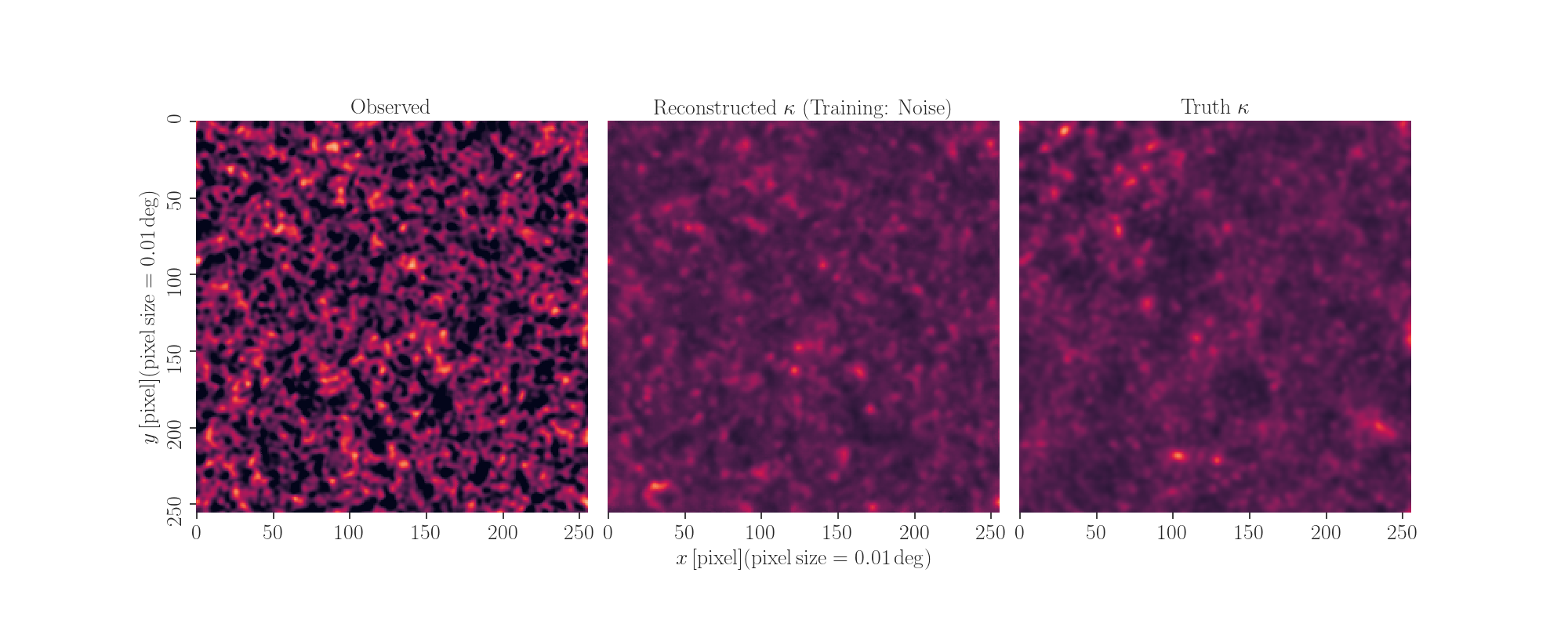}
     \caption{
     An example of image-to-image translation by our networks.
     The left panel shows input noisy convergence map, while the right is for true (noiseless)
     convergence. The medium represents the reconstructed map by our networks. For the reconstructed map,
     we first obtain the underlying noise field from the generator in our networks and then derive the convergence map
     by the residual between the input noisy map and the predicted noise.
     In this figure, we use the networks trained by 30000 image pairs.
     \label{fig:noise_train_image_fid}
  } 
    \end{center}
\end{figure*}

We examine three cases with 20000, 30000, and 40000 sets when training our networks.
After training, we input the test data set to the generator, in order to have lensing convergence predictions 
by the networks, denoted as $\kappa_{\rm DL}$. We use 1000 independent realizations of noisy lensing convergence  
for testing the network. Note that the data set in testing process is independent of the ones used in training.
We then measure the power spectrum and the PDF of $\kappa_{\rm DL}$. 
We also measure them for the corresponding true map $\kappa_{\rm true}$.
We normalize the lensing convergence so as to have a zero mean and a unit variance.
This normalization is  necessary to discern a small difference of summary statistics
between $\kappa_{\rm DL}$ and $\kappa_{\rm true}$.
We employ 20 bins logarithmically spaced 
in the range of $\ell=100$ to $10^5$ when measuring the power spectra.
We measure the PDF in 100 linear spaced bins in the range of $(\kappa-\mu)/\sigma = [-5, 15]$,
where $\mu$ is the mean and $\sigma$ is the standard deviation of lensing convergence over a sky coverage of 
$2.5\times2.5$ squared degrees.
Comparing the summary statistics of the predicted and true convergence fields, 
we will find how many realizations of input data sets are required to minimize the difference of statistics.
Furthermore, we study the dependence of model prediction by the networks on different realizations of training sets
by constructing 10 bootstrap subsamples of $20000-40000$ realizations from parent 60000 realizations.

\section{Results}\label{sec:results}

In this section, we summarize the performance of denoising weak lensing maps by our deep-learning approach.

\subsection{Visual impression}\label{subsec:visual}

We first show a visual comparison among three lensing fields, an input noisy convergence $\kappa_{\rm obs}$,
the predicted field by our networks $\kappa_{\rm DL}$, and the underlying noiseless convergence $\kappa_{\rm true}$.
Figure~\ref{fig:noise_train_image_fid} compares the lensing fields for a given realization.
The predicted convergence $\kappa_{\rm DL}$ is obtained by the networks trained with 30000 training sets.
As seen in the figure, the predicted field has a similar smoothness to the true $\kappa_{\rm true}$, 
and significant errors in the $\kappa_{\rm DL}$ field.
We also find some filamentary structures extending over a few degree lengths that look similar between $\kappa_{\rm DL}$ and $\kappa_{\rm true}$, 
while more compact structures such as 
peaks are found to be different.
Since such peaks in noiseless convergence are generated by massive galaxy clusters \cite{Hamana:2003ts},
the deep-learning denoising in this paper is probably less effective for searching for the massive objects in the Universe 
(we discuss this point in Section~\ref{subsec:pix2pix_comp} in more detail).
Nevertheless, the reconstructed convergence by the networks shows a similar level of clumpiness and morphology to the true convergence over arcmin to degree scales. From the comparison, we expect our networks can be good at estimating summary statistics 
including non-Gaussian information, but they may not work 
for reconstruction of highly compact and rare features in the image.

\subsection{Reconstruction of summary statistics}\label{subsec:avg_sum_stat}

\begin{figure*}
\begin{center}
       \includegraphics[clip, width=1.5\columnwidth, bb=60 10 908 432]
       {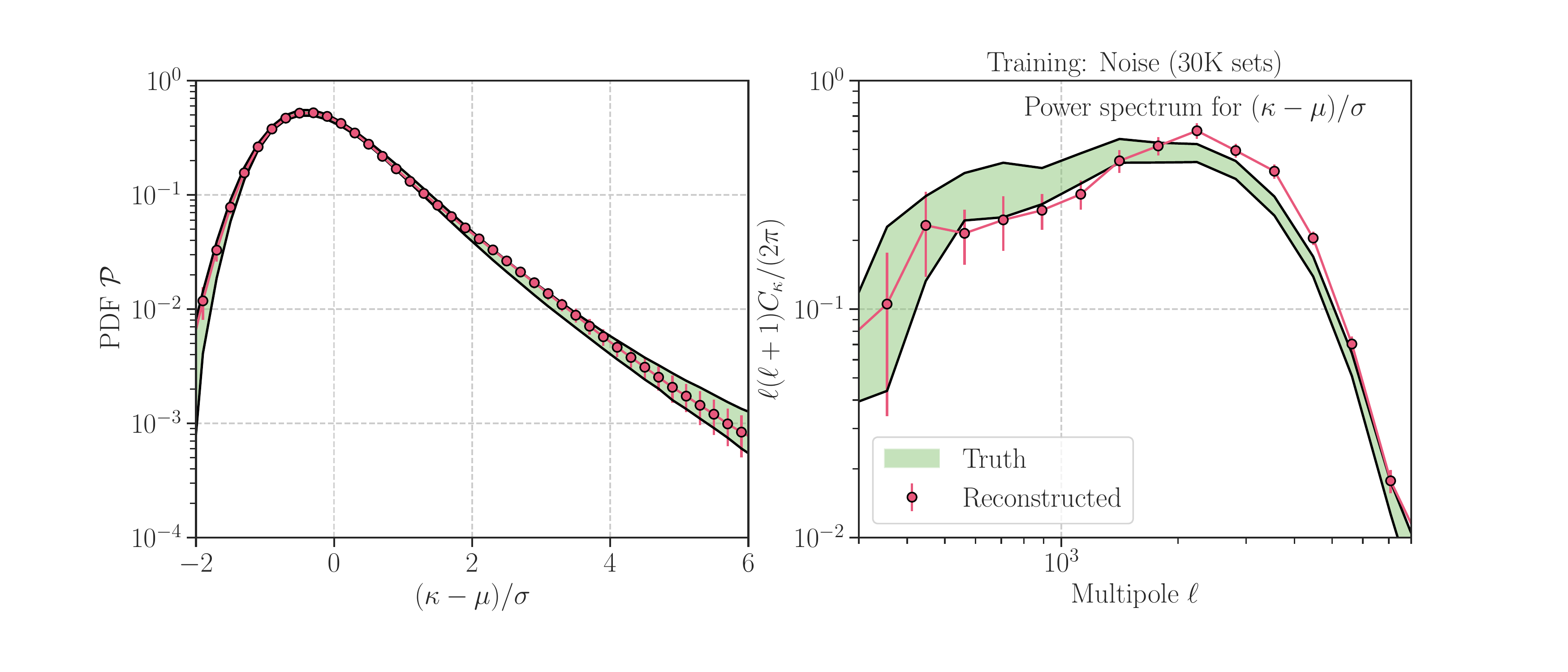}
     \caption{
     Comparison of weak-lensing summary statistics.
    The left panel shows the one-point probability distribution functions, and the right panel compares the power spectra.
    In each panel, we normalize the convergence so as to have zero mean and unit variance.
    The red points with error bars represent the summary statistics for the reconstructed field by our deep-learning method (the points and error bars are for the average and the rms over 1000 realizations for testing, respectively),
    while the green shaded regions show the $1\sigma$ variance of the true values. We evaluate the variance by using
    1000 noiseless weak lensing maps with a sky coverage of $2.5\times2.5$ squared degrees.
     \label{fig:summary_stats_30K}
  } 
    \end{center}
\end{figure*}

We discuss reconstruction of summary statistics.
Figure~\ref{fig:summary_stats_30K} shows the average of one-point PDFs and that of power spectra for the predicted field $\kappa_{\rm DL}$
over 1000 realizations of input noisy $\kappa_{\rm obs}$. In this figure, the red points show the average statistics for $\kappa_{\rm DL}$ and the error bars represents the standard deviation over 1000 realizations. For comparison, we also show the 1$\sigma$
variance of statistics as green filled region in the figure. This confidence region is defined by
the rms of PDF and power spectrum for true underlying fields $\kappa_{\rm true}$ over 1000 realizations.
Note that the results shown in Figure~\ref{fig:summary_stats_30K} are obtained by one of our networks with 30000 training sets 
(we have 10 networks that are trained with varying the number of training sets based on bootstrap sampling).
Figure~\ref{fig:summary_stats_30K} demonstrates that our deep-learning method can reconstruct the shape of the one-point PDF of underlying convergence field on average, while the reconstructed power spectrum has a systematic bias.

\begin{figure}
\begin{center}
       \includegraphics[clip, width=0.95\columnwidth]
       {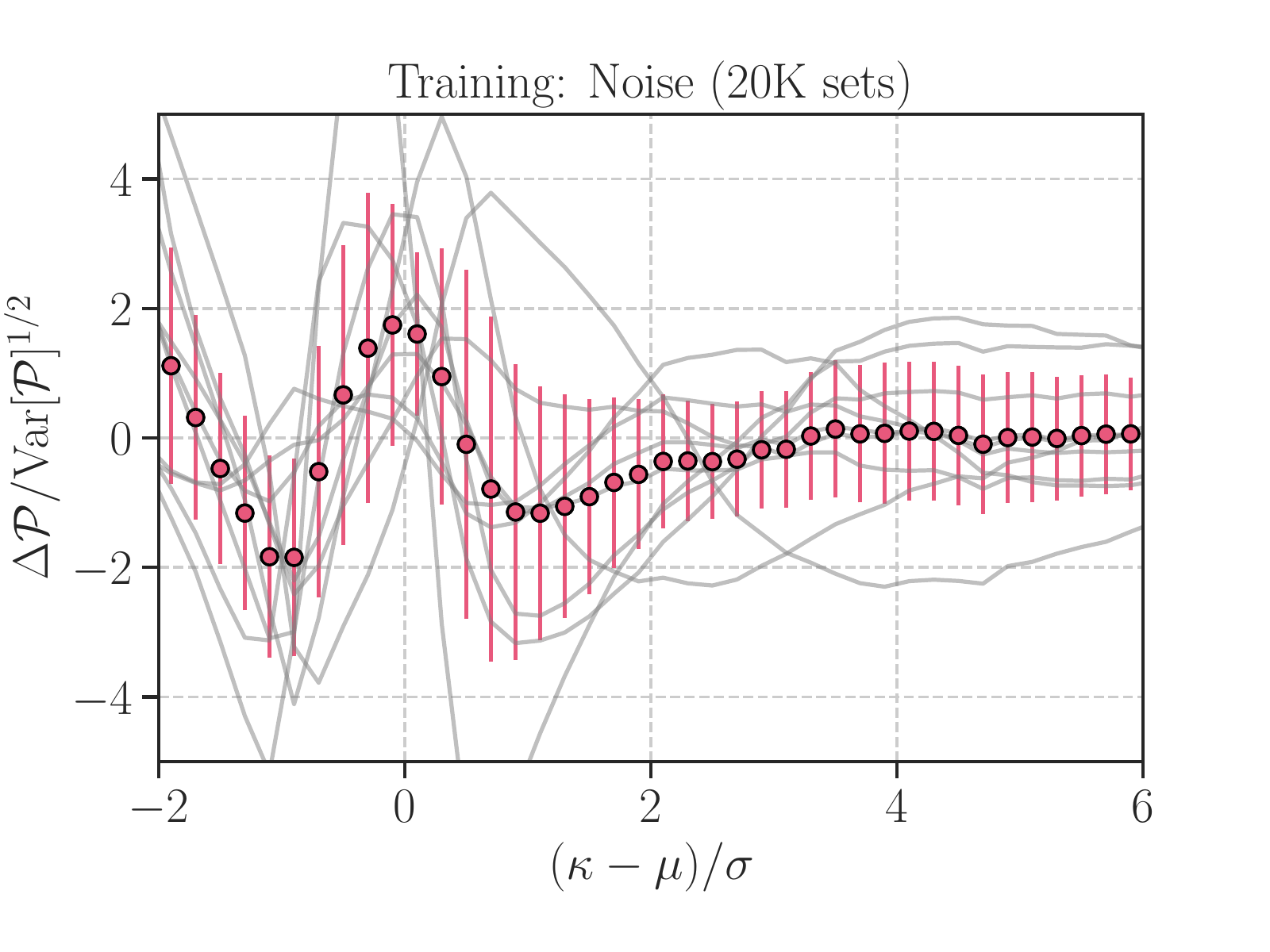}
       \includegraphics[clip, width=0.95\columnwidth]
       {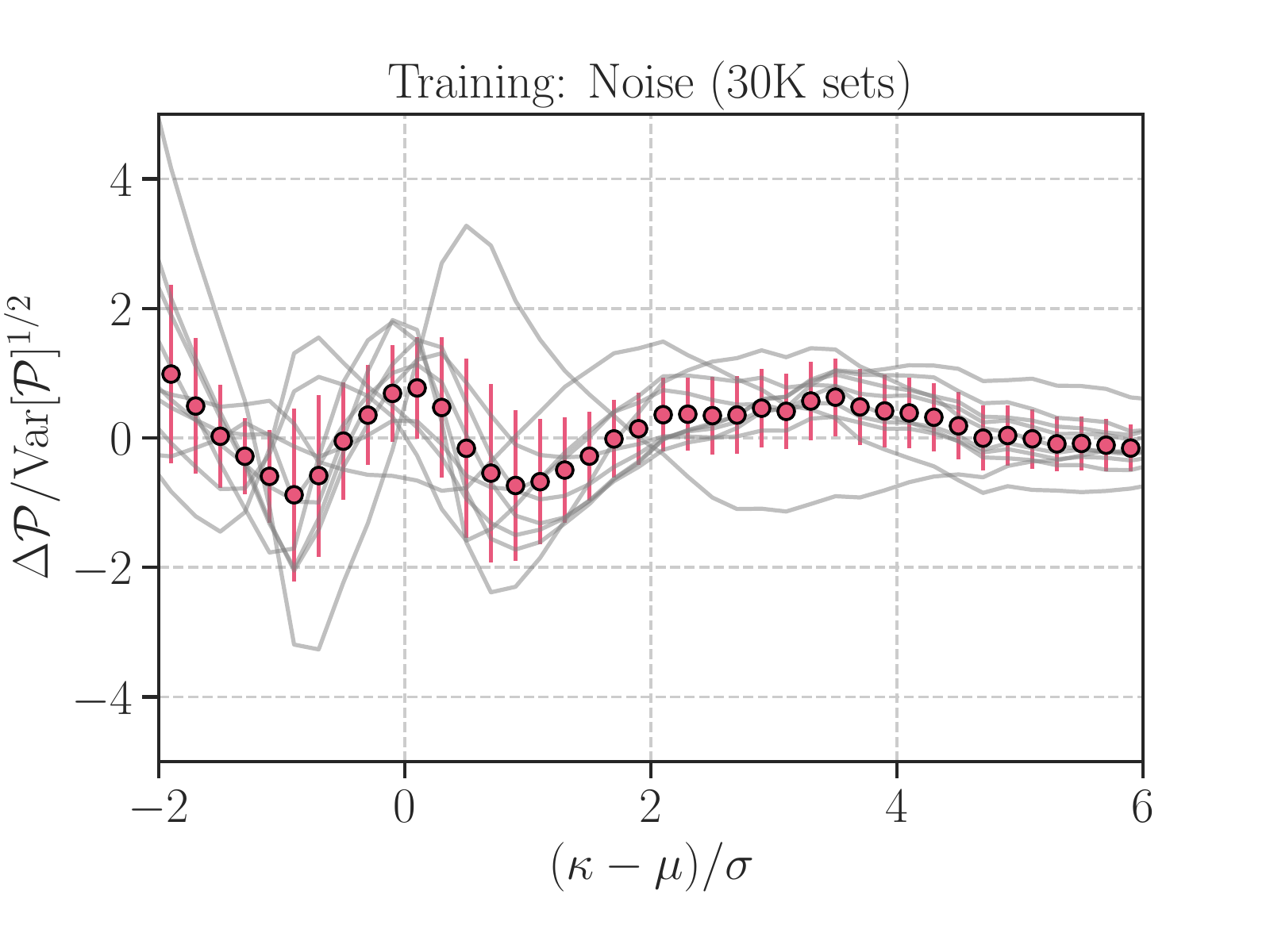}
       \includegraphics[clip, width=0.95\columnwidth]
       {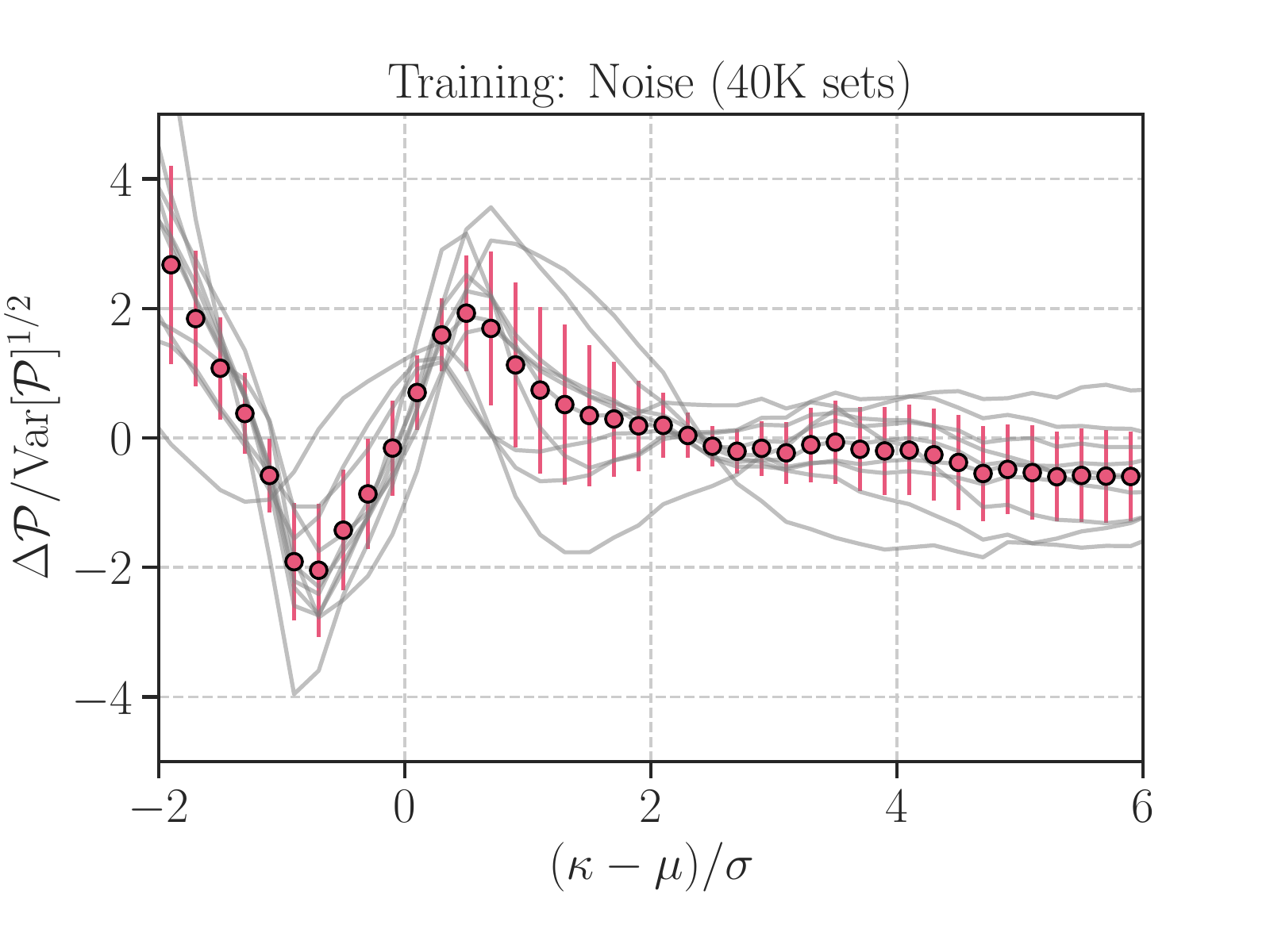}
     \caption{
     The dependence of the bias in one-point PDF of the convergence field on the number of training sets.
     Three different panels show the results by the network based on different numbers of training sets,
     20000, 30000, and 40000 image pairs from top to bottom.
     In each panel, the vertical axis shows the bias in convergence PDF with respect to the true noiseless counterpart
     and the horizontal axis is the value of convergence (normalized so as to have zero mean and unit variance).
     The gray lines show the results over 10 bootstrap realizations, while the red points and error bars are the average and standard deviation over 10 realizations.
     We note that the vertical axis is normalized by the statistical uncertainty of true PDF at each bin assuming a sky coverage of
     $2.5\times2.5$ squared degrees. The bootstrap scatter is found to be similar to the statistical uncertainty.
     \label{fig:pdf_bias_test}
  } 
    \end{center}
\end{figure}

We also examine the variation of the averaged PDF and that of power spectrum over 10 networks with different realizations.
Now we define the bias of a given statistic ${\cal S}$ at $i$-th bin as
\beq
{\rm Bias}[{\cal S}_{i}] \equiv \frac{\bar{\cal S}_{i}(\kappa_{\rm DL})-\bar{\cal S}_{i}(\kappa_{\rm true})}{{\rm Var}[{\cal S}_{i}(\kappa_{\rm true})]^{1/2}},
\eeq
where ${\cal S}_{i}$ is the statistic at $i$-th bin, 
$\bar{{\cal S}}_{i}(\kappa_{\rm DL})$ is the averaged statistic for the predicted field by our networks,
$\bar{{\cal S}}_{i}(\kappa_{\rm true})$ is the average for underlying noiseless field,
and ${\rm Var}[{\cal S}_{i}(\kappa_{\rm true})]$ represents the variance of ${\cal S}_{i}$ 
for noiseless field.

Figure~\ref{fig:pdf_bias_test} shows the bias in averaged PDF over 10 different networks.
The top, middle and bottom panels in the figure show the bias in PDF based on networks with 20000, 30000, and 40000 training sets, respectively.
In each panel, we show the result of 10 networks by bootstrap sampling of training sets.
The gray lines show the biases obtained by 10 different networks.
The red points and error bars are the average and rms over 10 realizations of network.
According to this figure, we find $\sim30000$ training sets are sub-optimal to have a less biased estimate of PDF on average.
The average bias level in reconstruction of PDF is found to be $\simlt1\sigma$ over the wide range of $|(\kappa-\mu)/\sigma|\simlt1$.
It would be worth noting that our networks will be able to reconstruct the PDF of noiseless lensing convergence at $(\kappa-\mu)/\sigma\sim0$ where the noise completely dominates in the original noisy map.
We also find the training of networks can depend on the realization of input data sets.
Even if we set the number of training data sets to be 30000, we find one of 10 networks shows a bias in PDF at 
$|(\kappa-\mu)/\sigma|\simlt1$ with a level of $\sim2\sigma$. Figure~\ref{fig:pdf_bias_test} clearly demonstrates that
the network will have a variety due to the statistical fluctuations in input data sets.
This suggests that training based on single realization of training sets is not sufficient for denoising weak lensing maps\footnote{In Appendix~\ref{app:var_kappa_after_denoise}, we examine the variation of the rms of PDF over 10 different realizations. We find that the variation coming from the network-wise fluctuation is less important in the rms of PDF for denoised maps.}.

\begin{figure}
\begin{center}
       \includegraphics[clip, width=0.95\columnwidth]
       {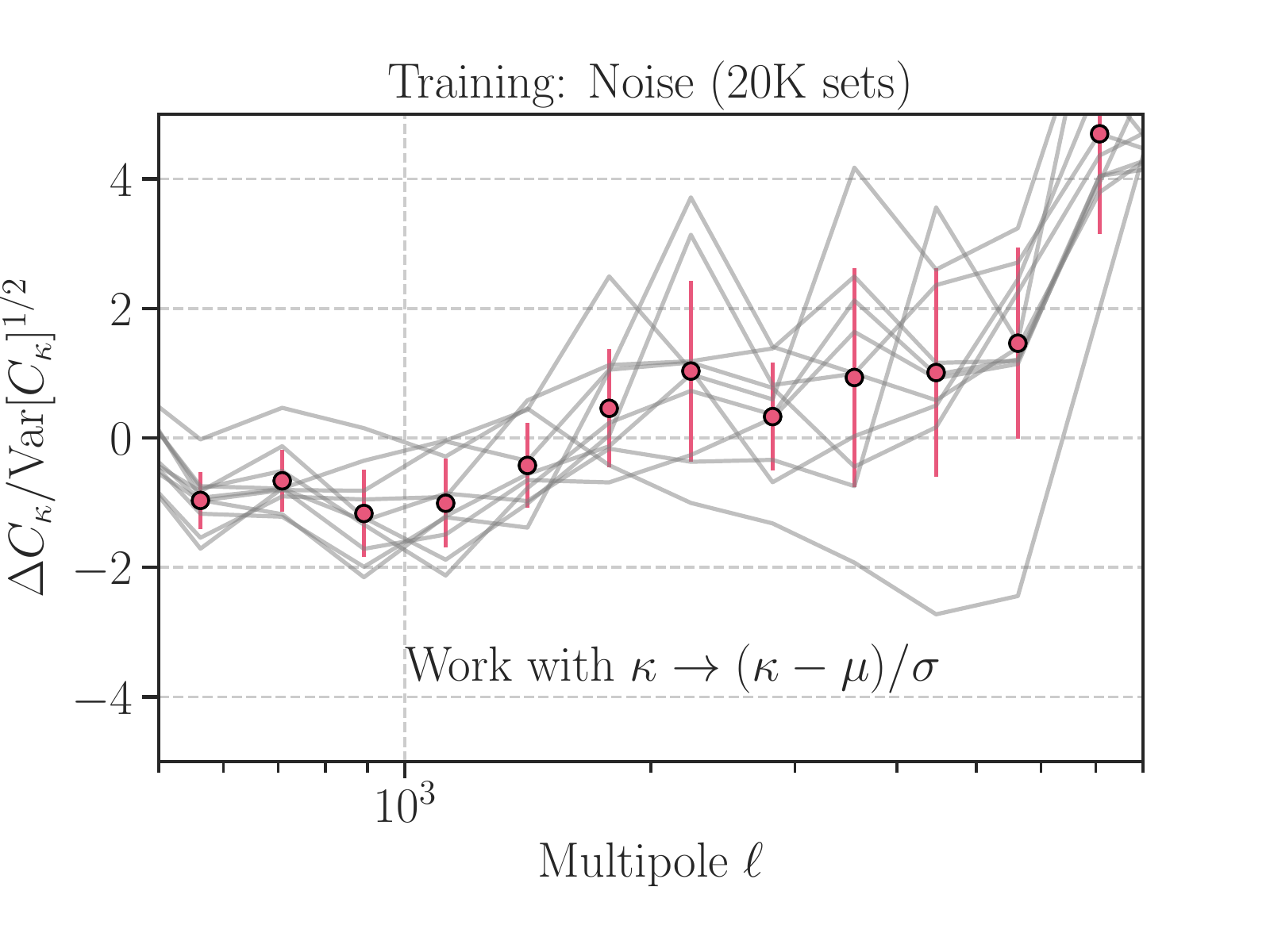}
       \includegraphics[clip, width=0.95\columnwidth]
       {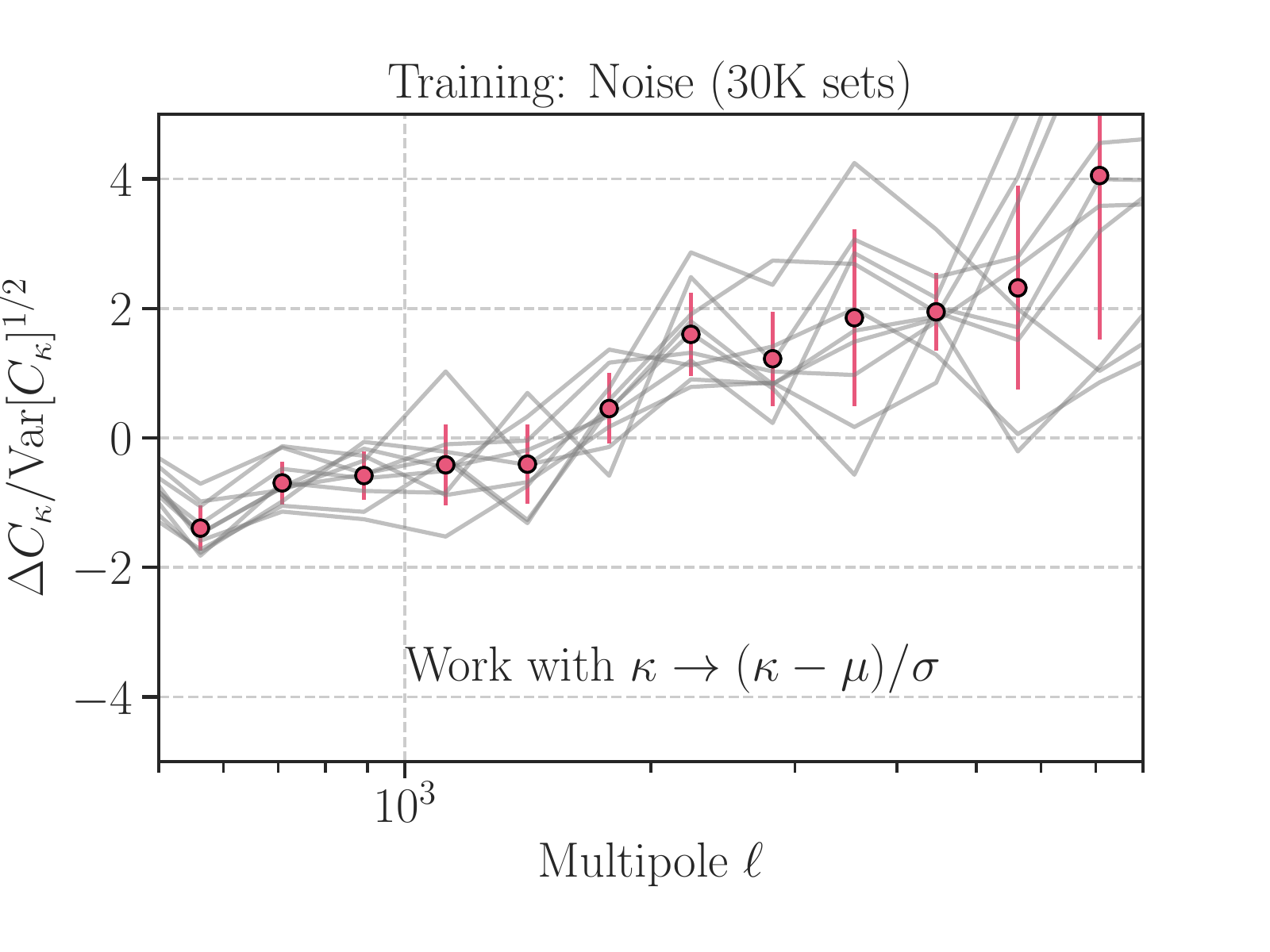}
       \includegraphics[clip, width=0.95\columnwidth]
       {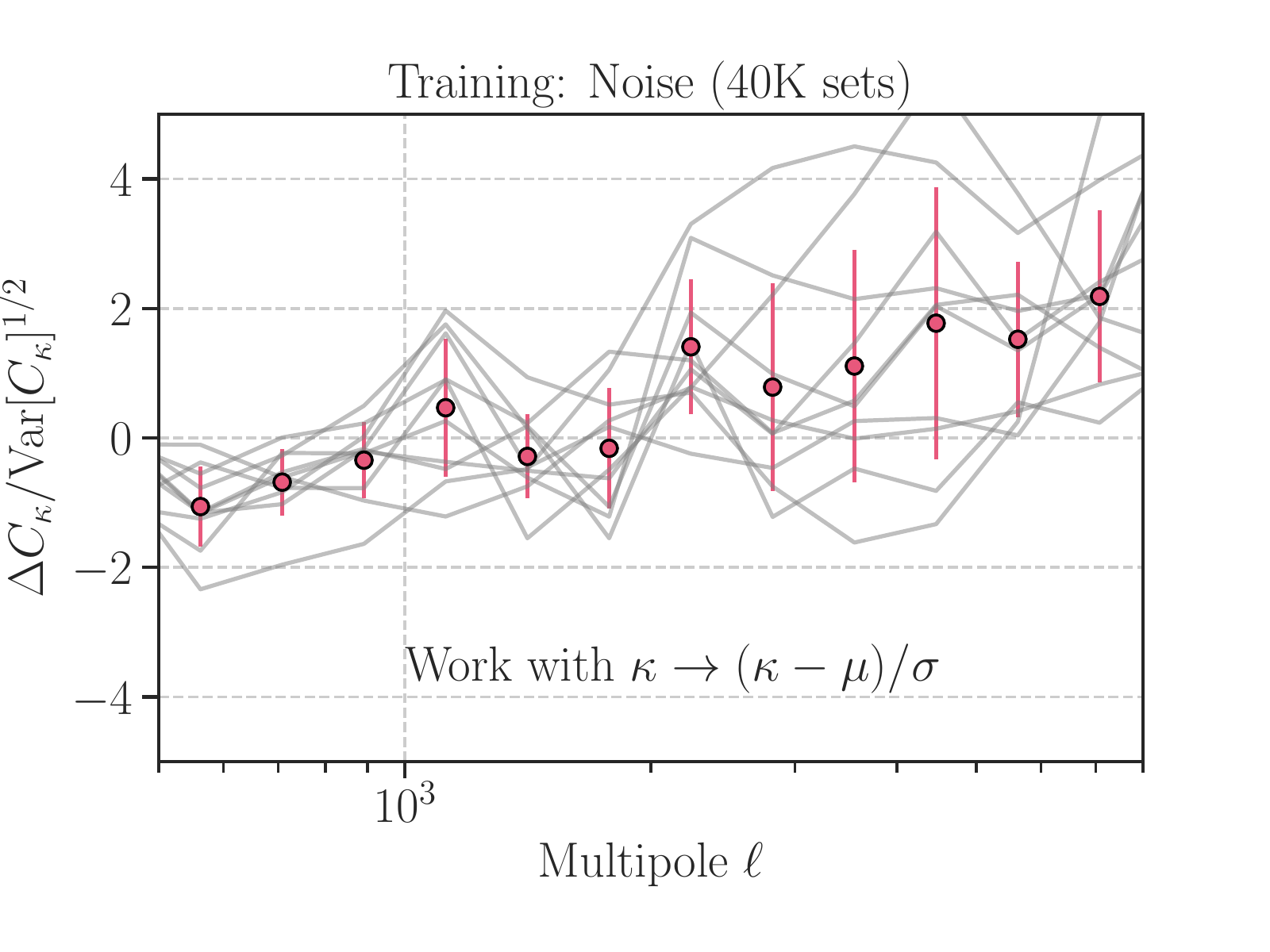}
     \caption{
     Similar to Figure~\ref{fig:pdf_bias_test}, but this figure is for the power spectrum.
     \label{fig:Cl_bias_test}
  } 
    \end{center}
\end{figure}

Figure~\ref{fig:Cl_bias_test} shows the bias in averaged power spectrum over 10 different networks.
Similar to the one-point PDF, we also find a non-negligible scatter over 10 different networks 
and $\sim40000$ realizations will be sub-optimal for an unbiased estimate of power spectrum.
Note that the lensing power spectrum can be estimated without denoising procedure and we already have some unbiased estimators
of lensing power spectrum in practice \cite{Hu:2000ax, 2011MNRAS.412...65H, 2016MNRAS.457..304B}. 
Therefore, we expect the biases shown in Figure~\ref{fig:Cl_bias_test} will not relevant to actual cosmological analyses. 
Rather, the figure indicates that our network may not use the Fourier decomposition to separate
the noise and underlying lensing fields. To see the reconstruction in Fourier space, we compute the cross correlation 
between the predicted and true noiseless maps. We find the ratio between the cross power spectrum and the true auto power spectrum
to be $\sim0.5$ at $\ell=500$ and $\sim0.1$ at $\ell=5000$. Hence, the denoising with our networks looks less efficient in Fourier space.

\subsection{Comparison on pixel-by-pixel basis}\label{subsec:pix2pix_comp}

\begin{figure}
\begin{center}
       \includegraphics[clip, width=0.85\columnwidth]
       {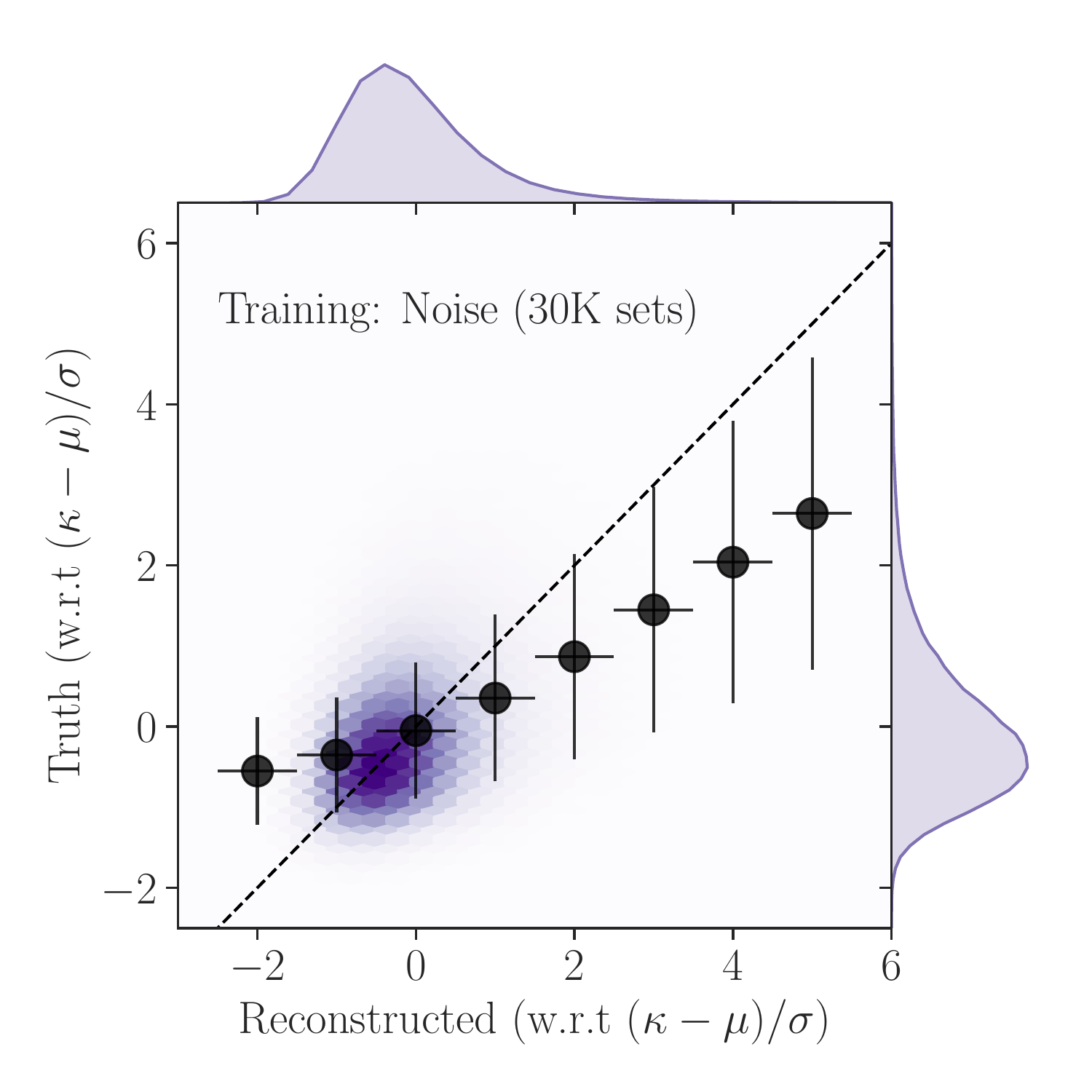}
     \caption{
     An example of scatter plot between the reconstructed convergence and true noiseless convergence.
     The dashed line shows the one-to-one correspondence. In this figure, we use the networks trained by 30000 image pairs
     and then adopt it to 1000 test data.
     \label{fig:scat_plot}
  } 
    \end{center}
\end{figure}

So far, we have focused on the statistical properties of weak lensing convergence predicted by our networks.
In this subsection, we test and validate our denoising method on a pixel-by-pixel basis.
Figure~\ref{fig:scat_plot} is s simple scatter plot of lensing convergence in $\kappa_{\rm DL}-\kappa_{\rm true}$ plane,
where $\kappa_{\rm DL}$ is the reconstructed one and $\kappa_{\rm true}$ is the noiseless counterpart.
For this figure, we show the results from 1000 realizations of $\kappa_{\rm DL}$ and $\kappa_{\rm true}$ from single networks trained 
by 30000 data sets. The dashed line in the figure represents the one-to-one correspondence, while the deeper color shows denser region
in the  $\kappa_{\rm DL}-\kappa_{\rm true}$ plane. 
In Figure~\ref{fig:scat_plot}, the black points with error represent the mean and standard deviation of $\kappa_{\rm true}$ 
for a given range of $\kappa_{\rm DL}$.
The figure shows that the predicted field by our networks follows some probability distribution with the width of $\sim1-3\sigma$ 
for a given $\kappa_{\rm true}$, where $\sigma$ is the rms of $\kappa_{\rm true}$ in a $2.5\times2.5$ squared-degrees sky, 
while the width slightly changes as the input value of $\kappa_{\rm true}$.
As seen in the figure, the mean relation in the $\kappa_{\rm DL}-\kappa_{\rm true}$ plane deviates from the one-to-one correspondence
at $(\kappa-\mu)/\sigma\simlt-1$ and $\simgt 2$.

In addition, 
we examined the corresponding 
positive peaks between noiseless and reconstructed (denoised) maps.
Note that previous numerical studies have shown that single massive galaxy clusters often present along the direction of peaks 
with a peak height larger than $\kappa/\sigma >3-4$ \cite{Hamana:2003ts}.
We first identify the local maxima in both of noiseless field $\kappa_{\rm true}$ and its denoised counterpart $\kappa_{\rm DL}$.
For selection of peaks, we set the threshold of peaks to be $(\kappa-\mu)/\sigma>3$. 
For a given position of lensing peak in $\kappa_{\rm DL}$, we search for the matched peaks in $\kappa_{\rm true}$ within a radius of 3 arcmin. This search radius is set to be larger than the smoothing scale but still smaller than the angular size of massive haloes at $z=0.1-0.7$ \cite{Hamana:2003ts}.
Among 1000 test data sets, we find 33563 and 44624 peaks in $\kappa_{\rm DL}$ and $\kappa_{\rm true}$, respectively.
The number of matched peaks is found to be 12,329. The corresponding completeness of lensing peaks in $\kappa_{\rm DL}$ 
is given by $12329/33553=36.7\%$ and the purity to be $12329/44624=27.6\%$. We confirmed these results are almost unaffected
even if the threshold of peaks is changed to
$(\kappa-\mu)/\sigma>4$, 5, or 6.
According to these results, our denoising method with CANs cannot always predict noiseless lensing convergence in unbiased way
and it would be less effective for collecting less massive galaxy clusters and/or completing a search of massive galaxy clusters in weak lensing
maps.

\section{Application to cosmological analysis}\label{sec:Fisher}

Although our denoising method does not perfectly reproduce the true
density distribution, it may still be used to  
estimate cosmological parameters accurately.
Figure~\ref{fig:scat_plot} clearly shows positive correlation between the denoised map $\kappa_{\rm DL}$ and its noiseless counterpart 
$\kappa_{\rm true}$. We thus expect that statistical analysis of denoised maps can be utilized for cosmological parameter inference.

To study quantitatively the possibility of extracting cosmological information from denoised lensing maps, we perform a Fisher analysis with one-point PDF.
We argue that the one-point PDF as a function of $(\kappa-\mu)/\sigma$ 
should contain 
information of non-Gaussianity.
For a given noisy lensing map, we generate 10 realizations of denoised lensing map 
from 10 different networks using 10 bootstrap sampling of training data sets. We then construct an estimator of the denoised one-point PDF as
\begin{equation}
\hat{{\cal P}}({\cal K}) = \frac{1}{N_{\rm CAN}}\sum_{i=1}^{N_{\rm CAN}} {\cal P}_{i}({\cal K}_{\rm DN}), \label{eq:noiseless_pdf_est}
\end{equation}
where ${\cal K} = (\kappa-\mu)/\sigma$ is the normalized lensing convergence so as to have zero mean and unit variance, 
$N_{\rm CAN}=10$ is the number of available networks, 
and ${\cal P}_{i}({\cal K}_{\rm DN})$ represents the one-point PDF of predicted lensing maps by $i$-th CAN.

\begin{figure}
\begin{center}
       \includegraphics[clip, width=1.1\columnwidth]
       {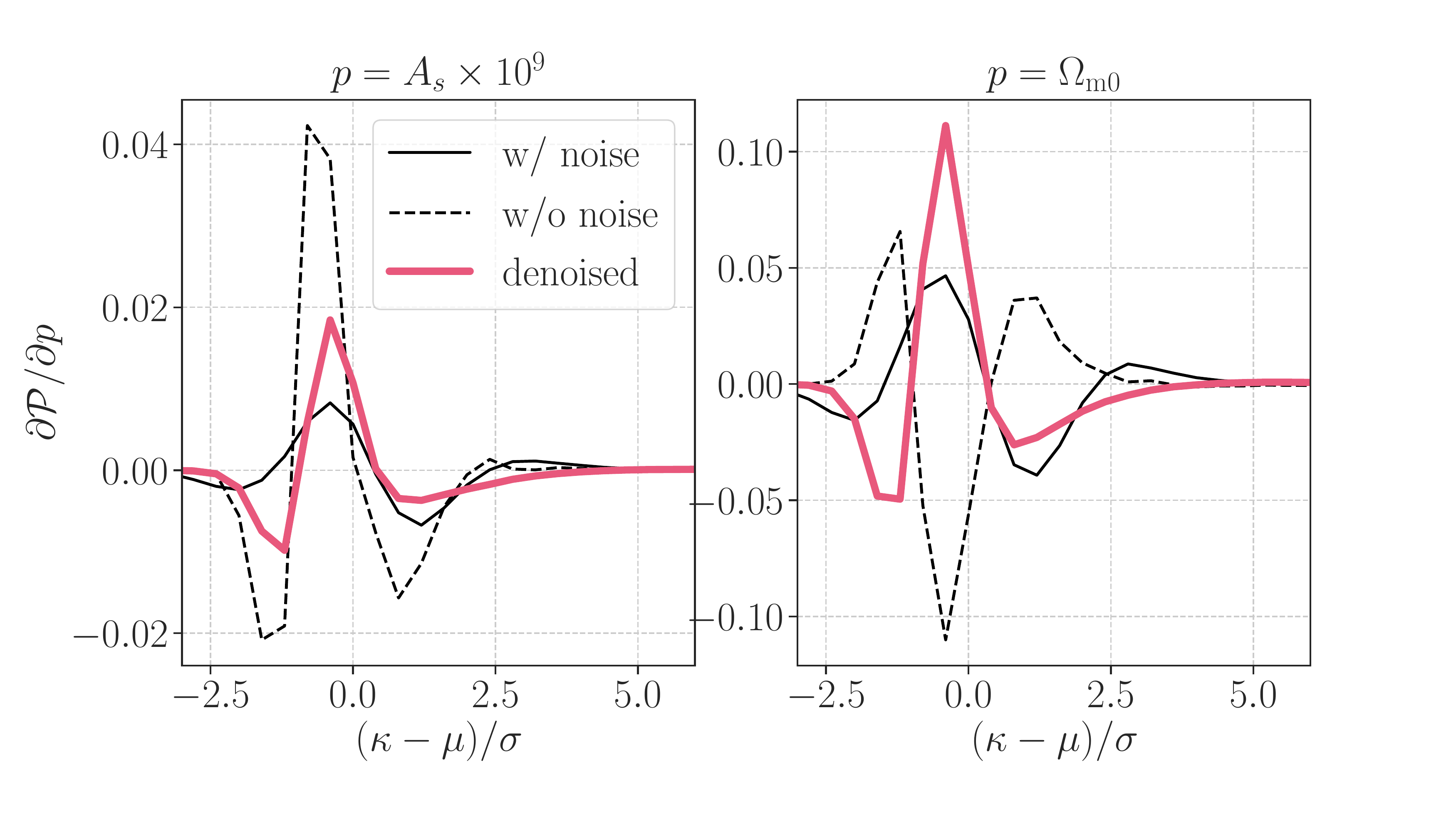}
     \caption{
     The parameter dependence of one-point PDF ${\cal P}$.
     Left and right panels show the first derivative of $\cal P$ with respect to the amplitude in curvature perturbations $A_{s}$
     and the mean matter density $\Omega_{\rm m0}$, respectively.
     In each panel, thick red line shows the parameter dependence of $\cal P$ for denoised lensing maps,
     while thin dashed and solid lines are for noiseless and noisy maps.
     We use the networks trained using 30000 lensing maps and perform denoising.
     Note that we normalized the field as $(\kappa-\mu)/\sigma$ on realization-by-realization basis
     and this procedure can make the cosmological dependence of statistics much complicated (see, e.g. Ref.~\cite{2013ApJ...774..111S}).
     \label{fig:pdf_cosmo}
  } 
    \end{center}
\end{figure}

We compute the Fisher matrix of the estimator given by Eq.~(\ref{eq:noiseless_pdf_est}) as
\beqa
F_{ij} &=& {\rm Tr}\left[ C^{-1} \left(\frac{\partial {\cal P}}{\partial p_{i}}\right) \left(\frac{\partial {\cal P}}{\partial p_{j}}\right)\right],
\label{eq:fisher}
\eeqa
where $C$ is the covariance matrix, ${\cal P}$ is the assumed model of one-point PDF, and  $p_i$ describes the parameters of interest.
The Fisher matrix provides an estimate of the error covariance for two parameters as 
$\langle \Delta p_{\alpha} \Delta p_{\beta} \rangle = (F^{-1})_{\alpha \beta}$,
where $\Delta p_{\alpha}$ represents the statistical uncertainty of parameter $p_{\alpha}$.
Here, we consider two cosmological parameters as an illustrative example: 
${\bd p}=(A_{s}\times10^9, \Omega_{\rm m0})$, where $A_s$ is the amplitude of the curvature perturbation
and $\Omega_{\rm m0}$ is the mean cosmic mass density.
We set the fiducial values to be $A_{s}\times10^9 = 2.35$ and $\Omega_{\rm m0} = 0.238$ in the Fisher analysis.
We calculate the cosmological dependence of one-point PDF $\cal P$,
by using 40 realizations of ray-tracing simulations with different cosmological models from our fiducial model \cite{2012ApJ...748...57S}.
We consider four different models by varying $A_{s} \times 10^{9}$ or $\Omega_{\rm m0}$ with $\pm10\%$ in logarithmic space.
These additional simulations are designed so as to cover the same sky coverage of $5\times5$ squared degrees with the source redshift of 1. We produce 1000 realizations of noisy convergence from the additional simulations in the same way as in Section~\ref{subsec:sims}.
Once we get the noisy convergence maps for the set of cosmological models, we compute the estimator of Eq.~(\ref{eq:noiseless_pdf_est})
from all noisy maps and evaluate the first derivative of one-point PDF with respect to the $i$-th parameter as
\beqa
\frac{\partial P({\cal K})}{\partial p_{i}} = 
\frac{\langle \hat{\cal P}({\cal K}; p_{i+}) \rangle - \langle \hat{\cal P}({\cal K}; p_{i-}) \rangle}{p_{i+} - p_{i-}}, \label{eq:deriv_P}
\eeqa
where $p_{i+}$ ($p_{i-}$) represents a higher (lower) parameter value than the fiducial one.
In Eq.~(\ref{eq:deriv_P}), $\langle \cdots \rangle$ is the average over 1000 realizations of noisy maps.
The data covariance in Eq.~(\ref{eq:fisher}) is also estimated from 1000 realizations of the estimator $\hat{\cal P}$ for our fiducial setup.
Because the covariance is estimated for a sky coverage of $2.5\times2.5$ square degrees,
we simply assume that the covariance matrix would scale with the inverse of sky coverage when making forecast for parameter constraints.
For comparison, we also compute the Fisher matrices of one-point PDF for noisy and noiseless lensing maps.


Before discussing the Fisher analysis result, we first look into the parameter dependence of $\cal P$ for denoised maps.
Figure~\ref{fig:pdf_cosmo} summarizes the parameter dependence evaluated by Eq.~(\ref{eq:deriv_P}).
There, the thick red lines show the parameter dependence of ${\cal P}$ for denoised maps. By comparing with the counterparts for noisy maps, we find that our denoising method with CANs 
can increase the sensitivity on the cosmological parameters at $(\kappa-\mu)/\sigma \sim 0$,
where the galaxy shape noise usually dominates in observations.
This highlights the 
possibility
of our denoising method to extract 
cosmological information from realistic observational data.

\begin{figure}
\begin{center}
       \includegraphics[clip, width=1.1\columnwidth]
       {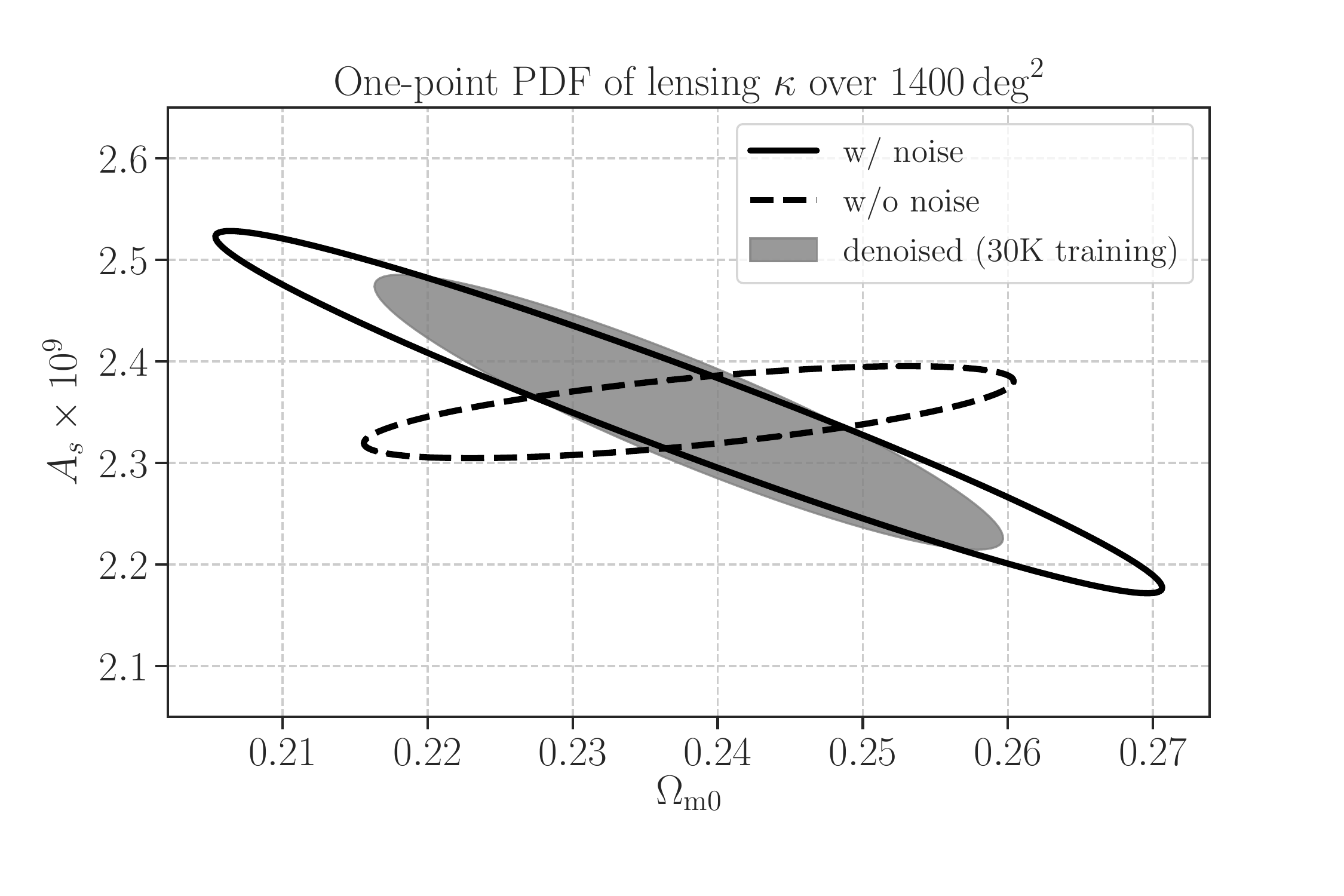}
     \caption{
     The expected $1\sigma$ constraints by one-point PDF of lensing maps 
     in a hypothetical galaxy imaging survey with a sky coverage of 1400 square degrees.
     To make a fair comparison, we restrict the range of $(\kappa-\mu)/\sigma$ in the Fisher analysis; we use only the one-point PDF of ${\cal P}$ greater than $0.01$.
     \label{fig:pdf_Fisher_HSC}
  } 
    \end{center}
\end{figure}

\begin{table}
\begin{center}
\scalebox{0.85}[0.85]{
\begin{tabular}{|c|c|c|}
\tableline
Lensing maps & 
$\Delta (A_{s}\times10^{9})$  &
$\Delta \Omega_{\rm m0}$  \\ \hline
With Denoising
& $8.96\times10^{-2}$ ($3.58\times10^{-2}$) & $1.43\times10^{-2}$ ($5.73\times10^{-3}$) \\
Noisy
& $1.18\times10^{-1}$ ($2.92\times10^{-2}$) & $2.16\times10^{-2}$ ($5.36\times10^{-3}$) \\
Noiseless
& $3.00\times10^{-2}$ ($2.22\times10^{-2}$) & $1.48\times10^{-2}$ ($1.09\times10^{-2}$) \\ \tableline
\end{tabular}
}
\caption{
\label{tab:fisher_err_HSC} 
Summary of the marginalized error of cosmological parameters by one-point PDF of lensing map.
Second and third columns list the expected statistical error assuming 1400 square degrees, 
and the number with bracket shows the un-marginalized error in the Fisher analysis. 
}
\end{center}
\end{table}

\begin{figure*}[!ht]
\begin{center}
       \includegraphics[clip, width=1.6\columnwidth, bb=60 80 1050 680]
       {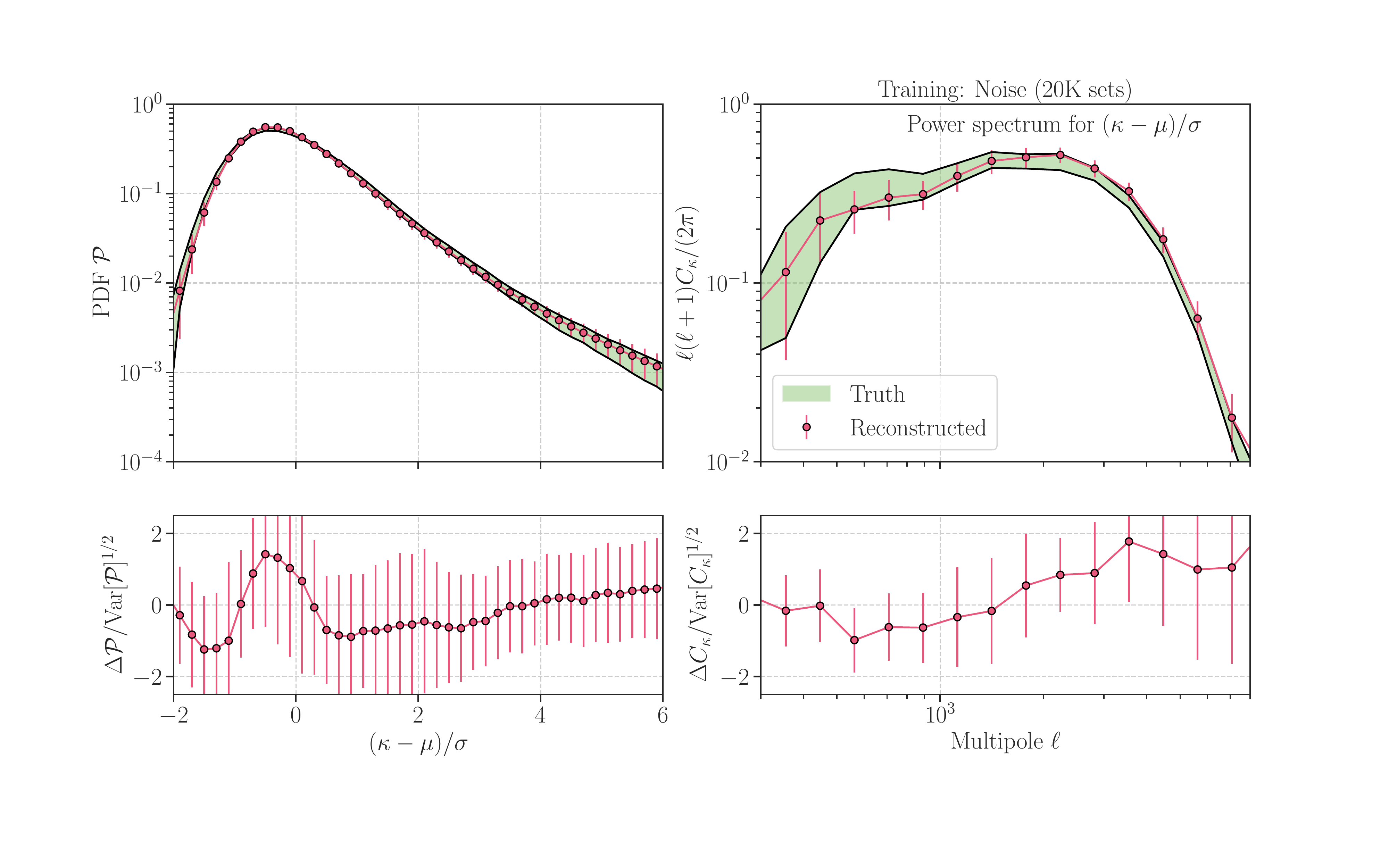}
     \caption{
     Performance for denoising in a hypothetical survey with the high source number density 
     of $40\, {\rm arcmin}^{-2}$. We plot the summary statistics as in Figure~\ref{fig:summary_stats_30K}. The bottom left panels show the average bias in the reconstructed PDF normalized to the rms of the noiseless PDF, while the bottom right panel is for the power spectrum. Note that the red error bars include the statistical error over 1000 test data as well as the bootstrap scatter over 10 different networks.
     \label{fig:40K_future}
  } 
    \end{center}
\end{figure*}

Figure~\ref{fig:pdf_Fisher_HSC} shows the expected parameter constraints 
for a sky coverage of 
1400 square degrees. To perform the Fisher analysis, we adjust the the range of $(\kappa-\mu)/\sigma$ in noisy
and denoised maps independently so that we can use the range with ${\cal P}\ge 0.01$. 
This threshold is found to correspond to $-2.5 < (\kappa-\mu)/\sigma < 2.5$ in noisy lensing maps,
and $-2 < (\kappa-\mu)/\sigma < 4$ in denoised and noiseless ones.
We find that the one-point PDF in denoised lensing maps can actually improve the constraints of $A_{s}$ and $\Omega_{\rm m0}$ by $30-40\%$, 
compared to the case using noisy lensing maps. We note that the improvement is not
significantly affected even if we vary the range of $(\kappa-\mu)/\sigma$ when
performing the Fisher analyses. Table~\ref{tab:fisher_err_HSC} summarizes the marginalized errors in cosmological parameters.

\section{Future prospects in upcoming survey}\label{sec:future}

Upcoming galaxy imaging surveys are aimed at generating accurate lensing maps
by increasing the source number density and the area coverage.
These include the Wide Field Infrared Survey Telescope (WFIRST\footnote{\url{https://wfirst.gsfc.nasa.gov/}}),
the Large Synoptic Survey Telescope (LSST\footnote{\url{https://www.lsst.org/}}),
and the Euclid satellite\footnote{\url{https://www.euclid-ec.org/}}.

Since the noise rms of weak lensing maps approximately scales with the inverse of the source galaxy number density,
we expect that the upcoming surveys with a higher source number density can provide less noisy maps.
We here examine if our denoising method works for such 
accurate observations in the near future.
As a representative example, we assume the source number density to be $40\,{\rm arcmin}^{-2}$,
and fix the rms of shape noise for individual galaxies $\sigma_{\epsilon}=0.35$ and source redshift $z_{\rm source}=1$.
In this case, the rms of noisy and noiseless map are found to be $1.7\times10^{-2}$ and $8.6\times10^{-3}$, respectively.
This set up roughly corresponds to the aforementioned upcoming imaging surveys.
Following the method in Section~\ref{subsec:sims} and setting $n_{\rm gal} = 40\,{\rm arcmin}^{-2}$ in Eq.~(\ref{eq:noise}), 
we produce new 60000 and 1000 realizations of input data set for training and testing networks, respectively. 
We then train the networks with 20000 realizations of these new data sets.
We obtain 10 different networks for denoising using 10 bootstrap sampling in total.

Figure~\ref{fig:40K_future} summarizes the performance of our networks for denoising.
In this figure, left and middle panels show the comparison of two summary statistics, one-point PDF and power spectrum.
The red points represent the average statistics over 10 bootstrap realizations and 1000 test data sets,
while the error bars are defined as
\beq
{\rm Var}[{\cal S}(\kappa_{\rm DL})]_{\rm stat} + {\rm Var}[\bar{{\cal S}}(\kappa_{\rm DL})]_{\rm bootstrap}.
\eeq
Here the first term is the standard deviation of statistics $\cal S$ of reconstructed $\kappa_{\rm DL}$
over 1000 realizations of test data, and the second term is the scatter in the average $\bar{\cal S}$ over 10 bootstrap networks.
If we increase the source number density, our networks perform well for both of the one point PDF and the power spectrum.

As in Section~\ref{sec:Fisher}, we perform the Fisher analyses of one-point PDF assuming  
the source number density of $40\,{\rm arcmin}^{-2}$ and the sky coverage of 20000 square degrees.
For this observational spec, our denoising method allows us to constrain $A_{s}\times10^{9}$ and $\Omega_{\rm m0}$ 
with a marginalized fractional error of $0.67\%$ and $0.95\%$,
while the PDF of noisy maps can provide the constraints of $A_{s}\times10^{9}$ and $\Omega_{\rm m0}$ with $0.85\%$ and $1.47\%$.
Note that the improvement of cosmological parameters by denoising is equivalent to the increase in sky coverage by $\sim50\%$ in the analysis of noisy lensing map.

\section{Conclusion and discussion}\label{sec:con}

We have studied denoising weak lensing maps 
using the conditional adversarial networks (CANs) developed in Ref~\cite{2016arXiv161107004I}.
We have developed a training strategy for the networks to denoise noisy lensing maps.
Our findings are summarized as follows.
The networks learn efficiently mapping from input noisy maps to the underlying noise field.
About 30000 realizations of image pairs are found to be sub-optimal for efficient training of networks to denoise weak lensing 
with a sky coverage of $2.5\times2.5$ squared degrees and with a similar noise level to the ongoing ground-base galaxy surveys \cite{Mandelbaum:2017dvy}. 
The trained networks reproduce the one-point PDF of noiseless weak lensing maps
with deviations within a $1\sigma$ level. 
Although the networks do not reproduce the noiseless power spectrum equally well
if we assume the noise level in the ongoing survey, 
we find that the performance can be improved in upcoming survey with higher source number density.
When assuming the source number density of $40\, {\rm arcmin}^{-2}$, the one-point PDF
and power spectrum are in good agreement with the true counterparts. 

We have studied reconstruction of noiseless lensing maps on a pixel-by-pixel basis 
and have found that denoising with CANs produces a biased estimate of local convergence.
Nevertheless, we find that a clear positive correlation between noiseless and denoised maps remains. We thus argue
that cosmological information imprinted in noiseless maps can be extratced at least partially with our denoising method.
We have demonstrated that our deep-learning denoising can improve the cosmological constraints
by increasing the sensitivity of one-point PDF to a few
cosmological parameters.
Assuming that the noise level is similar to the ongoing ground-base galaxy survey by Subaru Hyper-Suprime-Cam, 
we find the improvement by denoising can reach the level of $30-40\%$ in two primary cosmological parameters.
Our denoising approach enables us to tighten cosmological constraints by one-point PDF using data from upcoming galaxy imaging surveys.
We emphasize that the improvement by denoising is equivalent to the increase in the survey area by a factor of $1.5$.
We thus conclude that the denoising method developped in the present paper effectively maximizes the science output from ongoing and upcoming galaxy imaging surveys.

Denoising with deep-learning networks can be a powerful tool to probe the cosmic large-scale structure
with weak lensing.
It is yet to be investigated how the method works for real data sets.
In this paper, we have worked with an idealized situation while ignoring several effects in real observations.
These include inhomogeneous angular distribution of source galaxies, masking around bright stars,
photometric redshift uncertainty, biases in galaxy shape measurement, and the correlation between intrinsic ellipticity and lensing-induced shear (see, Ref~\cite{Mandelbaum:2017jpr} and 
the references therein for systematics in weak lensing measurement). 
Further studies are necessary to examine the applicability to real data sets.

In addition, more studies are desirable for reducing the bias in summary statistics induced by denoising.
As found in Section.~\ref{subsec:avg_sum_stat}, our denoising procedure cannot reproduce the two-point correlation of noiseless lensing field with a reasonable accuracy.
This would be partly because the objective of our networks does not include 
any information about the two-point correlation. It is interesting to develop
appropriate way providing the information of spatial correlations on the networks. 
We expect that there are still rooms for optimizing the training strategy for cosmological analysis.
As a brute-force approach, one can make use of a large set of numerical simulations for 
gravitational lensing and emulate the bias of statistics induced by denoising as a function of cosmological parameters. Although this is straightforward, one need reasonable prior information in the parameter space of interest in practice.

Nevertheless, 
the denoising with deep learning is a unique method at present to reconstruct the true weak lensing map
with an angular resolution of $\sim1\, \rm arcmin$. It is worth exploring this approach further for cosmological analyses. 
In principle, the image-to-image translation with deep learning can be applied 
for multi-dimensional data sets. Interesting examples in modern cosmology include reconstruction of the three-dimensional matter density field from the observed spatial distribution of galaxies and the component separation in intensity mapping observations over a wide range of frequencies of photons.
It is a challenging task to extract cosmological information in an efficient way from future cosmology surveys.
Deep leaning is among the promising approaches and is expected to play a crucial role in future statistical analyses.
Our method proposed in this paper may be one of the most potent forces 
that drive big-data analyses in the era of precision cosmology.

\acknowledgements
The authors appreciate careful reading and suggestion to improve the article by the anonymous referee.
The authors also thank Akisato Kimura for helpful comments.
This work was in part supported by 
Grant-in-Aid for Scientific Research on Innovative Areas
from the MEXT KAKENHI Grant Number (18H04358) and by JST CREST Grant Number JPMJCR1414.
Numerical computations presented in this paper were in part carried out 
on the general-purpose PC farm at Center for Computational Astrophysics, CfCA, 
of National Astronomical Observatory of Japan.

\appendix

\begin{figure*}[!ht]
\begin{center}
       \includegraphics[clip, width=2.2\columnwidth, viewport=100 50 1440 576]
       {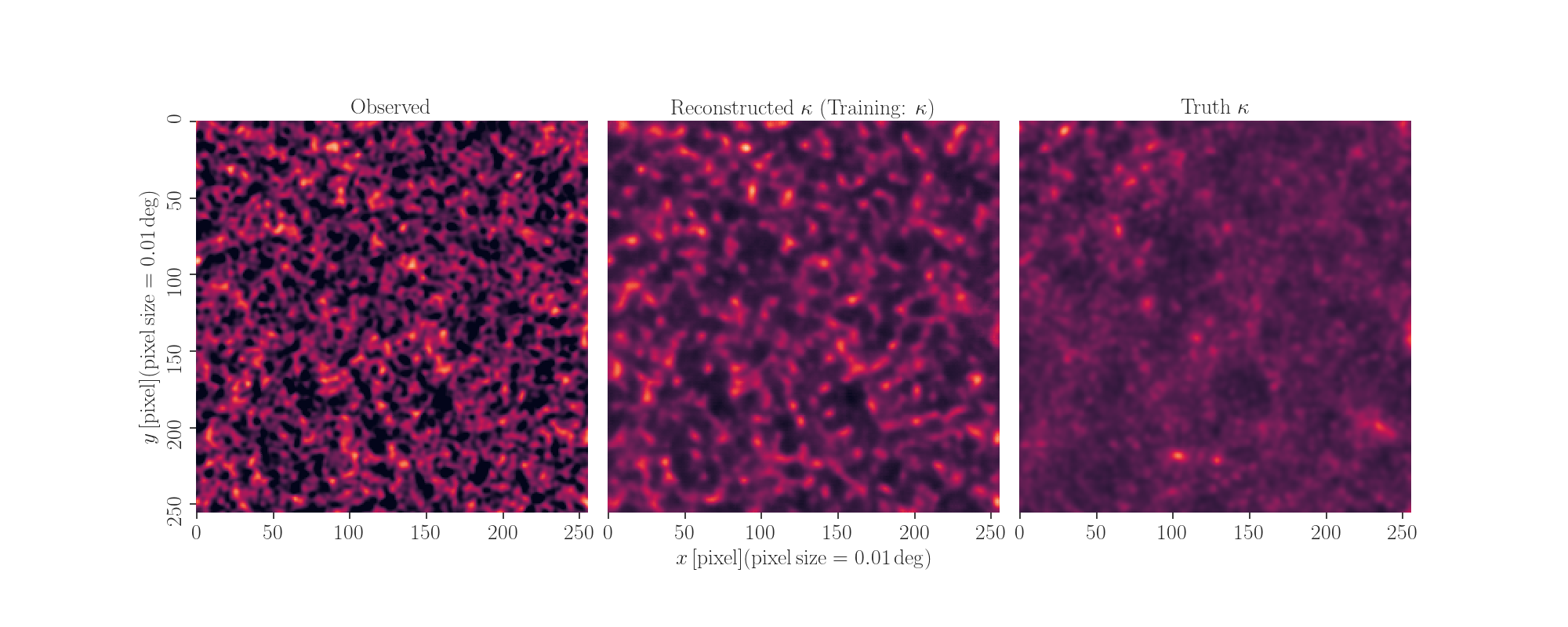}
     \caption{
        Similar to Figure~\ref{fig:noise_train_image_fid}, but the networks have been trained so that the generator can predict
        the lensing convergence.
        \label{fig:kappa_train_image}
     }
    \end{center}
\end{figure*}

\begin{figure}
\begin{center}
       \includegraphics[clip, width=0.9\columnwidth]
       {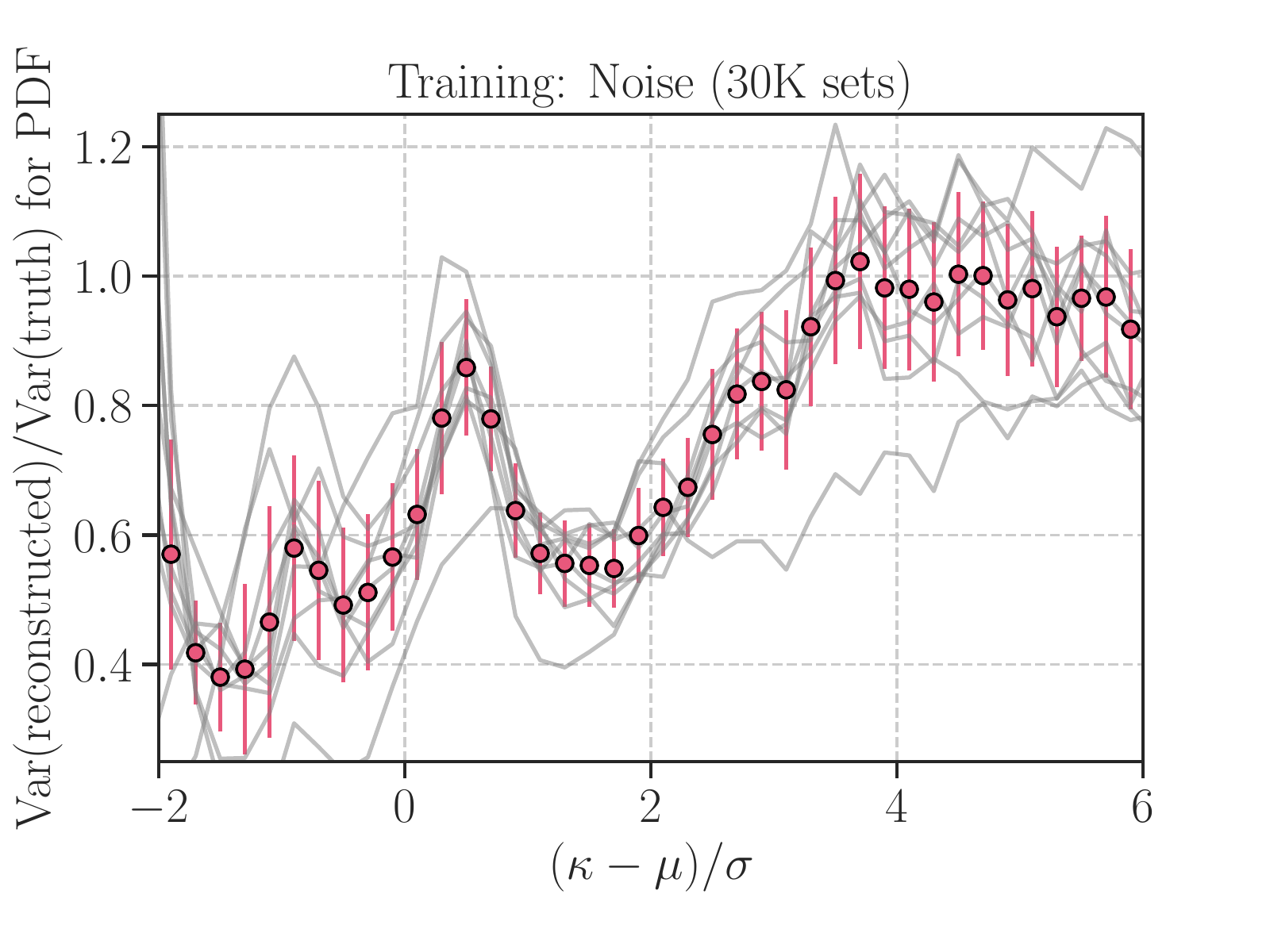}
     \caption{
         The scatter of one-point PDF of denoised map divided by 
         one for the noiseless counterpart. The gray lines represent the results over 10 bootstrap realizations of training datasets, while the red points and error bars are the average and standard deviation over 10 realizations.
        \label{fig:var_kappa_train}
     }
    \end{center}
\end{figure}

\section{Training to generate lensing convergence}\label{app:train_kappa}

In this appendix, we study another way of training the network for denoising weak lensing map.
We train the networks so that the generator will produce the underlying lensing convergence from an input noisy map.
Similar to the method in Section~\ref{subsec:training}, the training is most efficient when noisy data sets 
consist of $20000-40000$ independent realizations of noiseless convergence maps combined with 100 noise realizations. 
After some trials, we find that the summary statistics predicted by the networks do not agree with
the true counterparts within a $\sim3\sigma$ level. 

Figure~\ref{fig:kappa_train_image} shows the comparison between the predicted convergence by the networks 
and the true noiseless map. Apparently the networks fail to reproduce the smoothness of convergence 
with the same level as the truth, and there are too many clumpy structures in the predicted convergence.

\section{Variance of lensing statistics after denoising}\label{app:var_kappa_after_denoise}

We here study the variance in two summary statistics of reconstructed convergence.
Figure~\ref{fig:var_kappa_train} shows the scatter of one-point PDF
for reconstructed fields, divided by the scatter for noise-free lensing maps.
For this figure, we used the networks trained by 30000 image pairs and adopted them 
to 1000 realizations of test dataset to produce denoised maps. 
We then evaluated the variance of one-point PDF over 1000 denoised maps as well as noiseless true maps.
The gray line in the figure represents the scatter in variance for 10 different bootstrap realizations of deep-learning networks. \change{In practice, the scatter in variance of reconstructed PDF (shown in the red error bar) can be reduced when one work with the average PDF over bootstrap realizations of networks.}

We find the variance in one-point PDF for reconstructed field is close to that 
for true noiseless field at $(\kappa-\mu)/\sigma > 3$, but the variance for PDF of $\kappa_{\rm DL}$ can be smaller by a factor of $0.4-0.8$ at $(\kappa-\mu)/\sigma \simlt 3$.
\change{This underestimation in lower $(\kappa-\mu)/\sigma$ bins can be mitigated when we decrease the number of training sets. This implies that the variance of reconstructed PDF is subject to overfitting in our networks.}
Also, the variance in power spectrum for $\kappa_{\rm DL}$ is found to range from $75-100\%$ of the true variance.

\bibliography{ref_prd}

\end{document}